\documentclass[sigconf,nonacm,balance=false]{acmart}

\usepackage{booktabs}
\usepackage{graphicx}
\usepackage{mathtools}
\usepackage{multicol}
\usepackage{tikz}
\usetikzlibrary{positioning,arrows.meta,fit,backgrounds}
\usepackage{placeins}
\graphicspath{{figures/}}
\renewcommand\footnotetextcopyrightpermission[1]{}
\newcommand{\auto}{\texttt{bridge\_auto}}
\newcommand{\fixedbridge}{\texttt{bridge}}

\newcommand{\MAPE}{\operatorname{MAPE}}
\newcommand{\Normal}{\mathcal{N}}

\begin{document}

\title{Reconstructing Missing Mandi Prices in North-East India with Regional Factors}
\author{Pritam Deka}
\affiliation{\institution{Khyontek AI}\city{Guwahati}\country{India}}
\email{pritam@khyontekai.com}

\begin{abstract}
Agricultural price panels contain prolonged and clustered reporting gaps, yet their reconstruction is often evaluated by deleting isolated observations. We investigate a regional-factor approach to weekly mandi-price reconstruction in five North-East Indian states, with Bihar and Goa as additional comparison states. The model combines a common price path with persistent market-specific deviations; an adaptive ensemble weights alternative borrowing structures using internally masked reports. On a matched retrospective benchmark against ten baselines, pooled North-East mean absolute percentage error is 19.3\%, compared with 26.5\% for SAITS. The corresponding calendar-aligned panel/time-block bootstrap interval for the error difference is 2.8--12.7 percentage points. Removing the target state's monthly aggregate leaves pooled error at 19.4\%, but eliminates the apparent advantage in Goa. An origin-restricted pipeline combining week-close input processing with revised ensemble validation yields 19.7\% error, compared with 30.8\% for carry-forward; this contrast does not isolate individual pipeline changes. Gains vary across states, and uncertainty intervals remain too narrow in several small panels. These findings support further study of structured retrospective reconstruction, not deployment claims: model development used earlier results, publication dates for monthly inputs are assumed, and confirmatory evaluation on untouched data remains outstanding.
\end{abstract}

\keywords{agricultural prices, missing data, time-series imputation, dynamic factor models, market-price reconstruction, Agmarknet}
\maketitle

\section{Introduction}

Market-price histories provide the empirical basis for comparing agricultural markets and developing price forecasts. Research on price information and rural-market performance motivates the value of reliable market data, but does not establish a benefit from imputation itself \cite{goyal2010}. In India, the Agricultural Marketing Information Network (Agmarknet) records prices reported by regulated wholesale markets, commonly called mandis \cite{agmarknet}. These records are valuable precisely because they describe local markets, but local series are not necessarily continuous. A missing observation can reflect an interruption in reporting, a market closure, or the absence of trade in a commodity \cite{ma2019}. Consequently, an imputed series must be distinguished from an administrative record of actual transactions.

This distinction matters in the North-East Indian panels studied here (Table~\ref{tab:datastats}). In the selected Assam panels, 69.3\% of market--commodity weeks have no report. Coverage is also low in Meghalaya and Nagaland, whereas Manipur is substantially denser. The problem is therefore heterogeneous: some gaps separate otherwise regular reports, others extend across much of a market's history, and some coincide across markets. A method that interpolates a short interruption successfully may have little support for reconstructing a long absence. These figures describe the selected panels rather than all markets or commodities in the region.

Existing agricultural forecasting systems already address incomplete prices through interpolation, autoregressive models, and cross-market matrix completion \cite{ladhar2023,ma2019,bhardwaj2023}. Related food-price research treats the recovery of incomplete market surveys as a substantive modelling task \cite{andree2021}. These studies motivate the present work without resolving its central empirical question: under which reporting patterns does a structured cross-market model improve reconstruction in short, sparse weekly panels? We examine that question by evaluating against hidden real reports under clustered masks, sustained outages, and simpler random-deletion scenarios. We do not claim an exhaustive literature search or priority for agricultural price imputation.

Spatial price analysis motivates the possibility of shared commodity-price movements without identical local prices \cite{fackler2001,barrett2002}; factor models provide a statistical decomposition of common movement and series-specific variation \cite{stock2002,banbura2014}. Building on these ideas, we represent each market's log price by a regional factor and a persistent local residual. Conditional interpolation and forward propagation follow established state-space principles \cite{harvey1984,durbin2012}, while the importance of serially dependent residuals in economic-data imputation is emphasized by Ng and Scanlan \cite{ng2024}. Our particular implementation compares autoregressive and random-walk residual bridges, pooled helpers, and a drifting target--neighbour basis (Section~\ref{sec:method}). These are candidate modelling choices in this study, not established properties of every market. We estimate their weights by within-panel masking, using the general stacking approach \cite{wolpert1992,breiman1996}.

The information set is as important as the model. Smoothing conditions on later observations, whereas a current-price estimate must respect its evaluation cutoff and the release timing of its inputs \cite{durbin2012,giannone2008}. A second concern is indirect target leakage: if a retained monthly average incorporates reports subsequently hidden for evaluation, removing the weekly target cells alone does not remove their information. This conditional concern follows the general definition of target leakage \cite{kaufman2012,kapoor2023}; it is not a claim that the precise composition of Agmarknet's monthly aggregates has been independently verified. We therefore compare the retrospective benchmark with aggregate-exclusion and origin-restricted controls (Section~\ref{sec:controls}). Their measured effects are evidence from this study, while remaining information and design limitations are stated in Section~\ref{sec:limits}.

\begin{samepage}
The study makes three contributions:
\begin{itemize}
\item \textbf{A structured reconstruction method.} A compact regional-factor and residual-bridge ensemble for sparse mandi panels, with within-panel estimation of candidate weights (Section~\ref{sec:method}).
\item \textbf{A matched comparative evaluation.} A comparison with ten baselines across seven states, including gap-pattern, component, timing, and aggregate-exclusion analyses (Sections~\ref{sec:design} and~\ref{sec:results}).
\item \textbf{An assessment beyond point error.} Empirical interval-coverage and downstream sensitivity analyses that distinguish reconstruction accuracy from operational usefulness (Section~\ref{sec:results}).
\end{itemize}
\end{samepage}

On the common retrospective benchmark, pooled North-East MAPE is 19.3\%, compared with 26.5\% for SAITS. This is a result of the present study, not a general performance guarantee: simple methods remain competitive in some states, intervals under-cover, and earlier results informed model development. Untouched prospective evaluation remains necessary.

\section{Related Work}

\subsection{Agricultural price reconstruction and market dependence}

Missing observations are an established concern in agricultural market-information systems. Ma et al.\ use SoftImpute to complete sparse Agmarknet data before producing interpretable forecasts \cite{ma2019}. Ladhar et al.\ distinguish interpolation with access to later observations from a past-only setting, for which they use ARIMA \cite{ladhar2023}. Bhardwaj et al.\ develop a crop-price forecasting approach based on spatial and temporal learning \cite{bhardwaj2023}. Together with wholesale-price forecasting studies such as Jha and Sinha \cite{jha2014}, this literature establishes the practical context for evaluating the input price history separately from the downstream forecaster. Outside Agmarknet, Andr{\'e}e models incomplete food-price surveys to estimate subnational price inflation \cite{andree2021}.

Cross-market borrowing has an economic motivation, but it requires qualification. Spatial price analysis distinguishes co-movement, integration, and equilibrium; related prices need not obey a constant difference or imply frictionless trade \cite{fackler2001,barrett2002}. Work on Indian food-grain markets likewise examines integration in a policy and institutional setting \cite{ghosh2011}. We use a common factor and an optional drifting basis as statistical representations of shared movement, not as tests of market integration or identified transport costs.

\subsection{Missing-data and state-space foundations}

The reporting mechanism determines what masked observations can establish about truly missing values \cite{rubin1976,schafer2002,little2019}. Hiding an observed price permits direct scoring, but it does not identify the price distribution during non-trading weeks or informative reporting failures. Empirical gap masks make the deletion patterns less artificial; they do not by themselves reproduce a missing-not-at-random mechanism.

State-space methods provide a natural distinction between filtering and smoothing in incomplete time series \cite{jones1980,harvey1984,kalman1960,rauch1965,durbin2012}. Dynamic factor methods extend this idea to many related series and irregular observation patterns \cite{stock2002,doz2012,banbura2014}. Factor dimension is itself a model-selection question \cite{bai2002}; our one-factor representation is a parsimonious design choice, not an estimated finding that a single factor is sufficient. Mixed-frequency estimation and nowcasting additionally require an explicit account of when each input becomes available \cite{chow1971,mariano2003,giannone2008}. Ng and Scanlan emphasize serial dependence in imputation of economic indicators \cite{ng2024}. Our contribution is an application-specific, approximate fitting procedure that combines these ideas with residual bridging and inner-mask selection; it is not a new general likelihood estimator for dynamic factor models.

\subsection{Machine-learning imputation and evaluation}

General approaches include matrix completion \cite{mazumder2010}, chained-equation and tree-based imputation \cite{vanbuuren2011,stekhoven2012}, and univariate interpolation or state-space methods \cite{moritz2017}. Neural methods exploit temporal and cross-variable dependence through recurrent architectures, including M-RNN and BRITS \cite{yoon2019,cao2018}, or self-attention, as in SAITS \cite{du2023saits}. Other approaches model missingness for downstream prediction \cite{che2018}, construct probabilistic imputations with variational or diffusion models \cite{fortuin2020,tashiro2021}, or adapt general time-series backbones \cite{wu2023}.

Recent alternatives represent spatial and temporal relations explicitly: GRIN uses graph-based recurrent imputation \cite{cini2022}, SPIN uses spatiotemporal attention \cite{marisca2022}, DeepMVI combines multiscale temporal and cross-series signals \cite{bansal2021}, and ImputeFormer introduces low-rank Transformer structure \cite{nie2024}. These methods were not run in our benchmark.

Published surveys and comparative studies show the breadth of available methods and the importance of spatiotemporal structure and evaluation settings \cite{wang2025,zahmatkesh2026,khayati2020}; PyPOTS provides implementations of several neural baselines \cite{du2023pypots}. We compare selected representatives under the present panel sizes and training budgets. Their results should not be generalized to all neural imputers or to alternatives trained on larger pooled datasets. The distinctive evaluation concern here is the combination of short histories, clustered reporting gaps, and potentially overlapping aggregate helpers.

Validation should reflect temporal, spatial and hierarchical dependence \cite{roberts2017}. Conventional cross-validation can be valid for some autoregressive models with uncorrelated errors \cite{bergmeir2018}, but this does not justify arbitrary random masking here. Preprocessing, overlapping inputs and model selection can introduce additional leakage \cite{kaufman2012,kapoor2023}.

\subsection{Ensembling and uncertainty assessment}

Stacking combines candidate predictions using held-out responses \cite{wolpert1992,breiman1996}. We use non-negative least squares followed by normalization, with the fitting responses drawn from internally masked reports. This implementation is related to, but not identical to, optimization directly over the probability simplex. Its weights describe a mixture of correlated predictors, not causal measures of component importance.

Predictive calibration also uses held-out errors, but formal coverage results depend on the construction and assumptions of the calibration procedure \cite{lei2018,gibbs2021}. Work beyond exchangeability provides tools for particular departures from standard assumptions, not an automatic guarantee for any dependent calibration sample \cite{barber2023}. Our intervals are rescaled using the same inner predictions that estimate ensemble weights. They are therefore empirical uncertainty summaries rather than split-conformal intervals with guaranteed coverage. The outer masked reports provide an additional coverage diagnostic, including evidence of failure in small panels and long gaps.

\section{Data}\label{sec:data}

\subsection{Source and study population}

The study uses locally downloaded responses from the Agmarknet 2.0 daily dashboard and monthly state-price endpoints \cite{agmarknet}. The daily loader retains rows marked reported with a positive modal price, and removes duplicates by date, market identifier, and commodity. It reads the price and arrival units supplied by the API. The downloads cover August 2024--October 2026 for Assam, Meghalaya, Manipur, Nagaland, Tripura, West Bengal, Bihar, and Goa; monthly state series extend back to 2021. West Bengal is a helper state rather than an evaluation state. Jharkhand yielded no daily rows in the local retrieval and contributes no daily market series.

For each evaluation state, commodities are ranked by the number of retained daily rows, with a minimum of 150 rows and a maximum of 20 commodities. Target markets require ten reported weeks for panel construction and twenty for masked evaluation. Daily helper markets require twenty reported weeks; monthly helper series require three reported months. Table~\ref{tab:datastats} describes these selected panels. These eligibility rules exclude the least observed series, and the evaluation does not quantify performance for markets with no historical reports.

\begin{table}[t]
\caption{Panels used in the study. Series are market-commodity pairs of the target state with at least ten reported weeks (20 best-reported commodities at most); observed share is the share of their weekly cells with a report.}
\label{tab:datastats}
\small
\begin{tabular}{lrrrrr}
\toprule
State & Reports & Markets & Comm. & Series & Observed (\%) \\
\midrule
Assam & 77{,}893 & 73 & 20 & 757 & 30.7 \\
Meghalaya & 7{,}021 & 17 & 12 & 106 & 25.6 \\
Manipur & 13{,}474 & 5 & 8 & 40 & 75.9 \\
Nagaland & 10{,}409 & 19 & 12 & 120 & 35.8 \\
Tripura & 97{,}219 & 32 & 16 & 330 & 60.4 \\
Bihar & 17{,}730 & 72 & 11 & 143 & 21.9 \\
Goa & 3{,}236 & 1 & 5 & 5 & 98.6 \\
\bottomrule
\end{tabular}
\end{table}

\subsection{Cleaning and weekly aggregation}

The retrospective cleaning rule flags a report when its price differs by a factor greater than three from both the same-market median in a centered 43-day window and the state--commodity median in the same week. It also flags deviations greater than a factor of eight from the full-sample state--commodity median. The procedure is intended to reduce isolated transcription or unit errors, but it does not establish that every flagged value is erroneous or that every retained price has a consistent unit. In particular, it is not a substitute for a complete unit-harmonization audit.

Within a market and commodity, daily price levels are averaged using non-negative arrivals plus a small numerical offset as weights, then transformed to logs. Weeks begin on Monday. A week with no retained report remains missing; it is not interpolated by the loader. The resulting matrix has weeks as rows and series as columns. Its first columns are the target state's markets, followed by daily helper markets and monthly state averages. In the original offline benchmark, monthly values are placed in the mid-month week, including the target state's own average. The timing and overlap implications are examined in Section~\ref{sec:controls}.

Table~\ref{tab:helpers} lists helper states. A listed state contributes only a monthly series when daily data were not downloaded. Geographic proximity motivates this selection, but trade-flow data were not used to verify the links. Panel lengths differ by commodity; the calendar date, not its relative row index, determines alignment in the dependence analysis.

\begin{table}[tbp]
\caption{Helper states for each target. Contributions depend on locally available daily and monthly data.}
\label{tab:helpers}
\small
\begin{tabular}{lp{5.1cm}}
\toprule
Target & Helper states \\
\midrule
Assam & West Bengal, Meghalaya, Bihar, Tripura, Nagaland, Manipur \\
Meghalaya & Assam, West Bengal \\
Manipur & Assam, Nagaland \\
Nagaland & Assam, Manipur \\
Tripura & Assam, West Bengal \\
Bihar & West Bengal; Jharkhand and Uttar Pradesh (monthly only) \\
Goa & Maharashtra and Karnataka (monthly only) \\
\bottomrule
\end{tabular}
\end{table}

\section{Regional-Factor Reconstruction}\label{sec:method}

\begin{figure*}[tbp]
\centering
\begin{tikzpicture}[
  box/.style={draw, rounded corners=2pt, align=center, font=\footnotesize, minimum height=8mm, fill=blue!6, inner sep=3pt},
  cand/.style={draw, rounded corners=2pt, align=center, font=\scriptsize, minimum height=5.2mm, minimum width=3.7cm, fill=orange!12, inner sep=2pt},
  outbox/.style={draw, rounded corners=2pt, align=center, font=\footnotesize, minimum height=8mm, fill=green!10, inner sep=3pt},
  arr/.style={-{Latex[length=2mm]}, thick},
  lab/.style={font=\scriptsize\itshape, align=center}]
\node[box, text width=2.9cm] (raw) at (0,0.95) {Agmarknet daily reports\\target and helper states};
\node[box, text width=2.9cm] (mon) at (0,-0.65) {Monthly state\\averages};
\node[box, text width=3.2cm] (panel) at (4.2,0.15) {Clean and aggregate\\Weekly log-price panel $Y$\\target markets $|$ helpers};
\draw[arr] (raw.east) -- ++(0.45,0) |- ([yshift=2.5mm]panel.west);
\draw[arr] (mon.east) -- ++(0.45,0) |- ([yshift=-2.5mm]panel.west);
\node[cand] (c1) at (8.6,1.9) {1: factor + AR bridge};
\node[cand] (c2) at (8.6,1.25) {2: factor + random-walk bridge};
\node[cand] (c3) at (8.6,0.6) {3: factor + drifting neighbour basis};
\node[cand] (c4) at (8.6,-0.05) {4: factor + pooled neighbours};
\node[cand] (c5) at (8.6,-0.7) {5: AR bridge only};
\node[cand] (c6) at (8.6,-1.35) {6: random-walk bridge only};
\begin{scope}[on background layer]
\node[draw, dashed, rounded corners=3pt, fit=(c1)(c2)(c3)(c4)(c5)(c6), inner sep=2.5mm, fill=orange!4] (group) {};
\end{scope}
\node[lab, above=0.5mm of group] {candidate fills of $Y$};
\draw[arr] (panel.east) -- (group.west |- panel.east);
\node[box, text width=3.0cm] (inner) at (13.0,1.35) {Inner masked reports\\retained after outer hiding\\three original masks};
\node[box, text width=3.0cm] (nnls) at (13.0,-0.45) {NNLS coefficients\\then normalize to weights\\Equations~\eqref{eq:nnls}--\eqref{eq:weights}};
\node[box, text width=3.0cm] (mix) at (13.0,-3.0) {Weighted log-price fill\\working variance includes\\candidate disagreement};
\draw[arr] (group.east |- inner.west) -- (inner.west);
\draw[arr] (inner) -- (nnls);
\draw[arr] (nnls) -- (mix);
\node[box, text width=3.2cm] (cal) at (8.6,-3.0) {Empirical variance scaling\\gap bins: 1, 2--3,\\4--8, 9+ weeks};
\node[outbox, text width=3.2cm] (res) at (4.2,-3.0) {Filled panel and\\nominal intervals\\not a coverage guarantee};
\node[outbox, text width=2.9cm] (use) at (0,-3.0) {Retrospective diagnostics\\and downstream\\sensitivity analyses};
\draw[arr] (mix) -- (cal);
\draw[arr] (cal) -- (res);
\draw[arr] (res) -- (use);
\end{tikzpicture}
\caption{Pipeline overview. Candidate predictions are combined using nonnegative coefficients normalized after fitting, followed by empirical interval scaling. Helper-dependent candidates are omitted without helpers. The information regime determines which inputs are allowed; control C also changes inner validation. Nominal intervals and downstream analyses do not establish deployment readiness.}
\Description{Daily and monthly reports form a weekly panel, six candidate families feed inner validation and normalized NNLS weights, and the weighted fill receives empirical variance scaling before retrospective evaluation.}
\label{fig:pipeline}
\end{figure*}
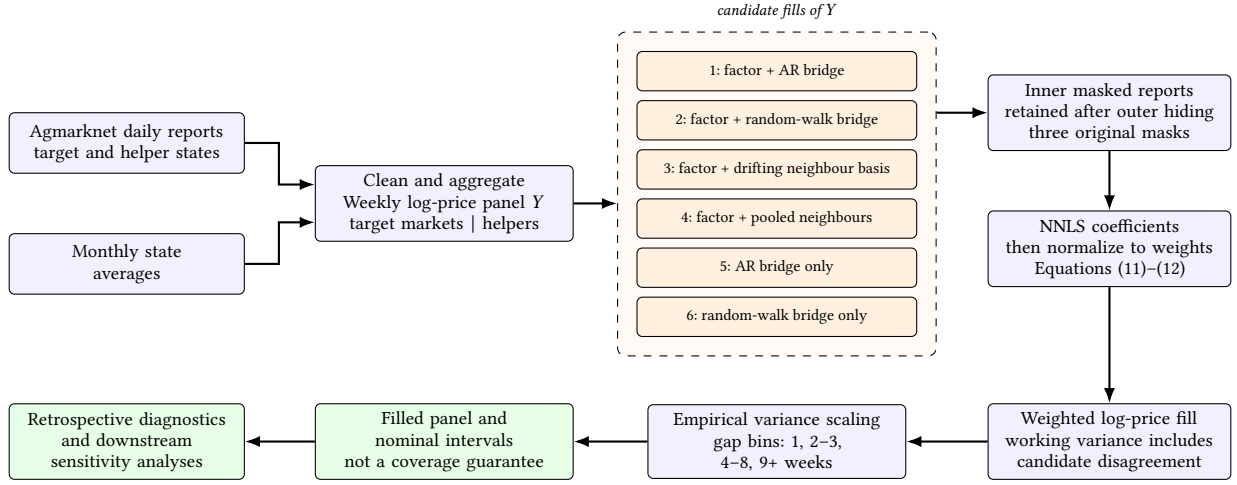

Figure~\ref{fig:pipeline} summarizes the reconstruction pipeline. The candidate family, normalized NNLS weights, and empirical interval scaling are described below; the diagram does not imply that nominal intervals have achieved their stated coverage.

\subsection{Task, notation, and information sets}

For a single commodity, let $p_{mt}>0$ denote a weekly price and $y_{mt}=\log p_{mt}$ its log, for series $m=1,\ldots,N$ and weeks $t=1,\ldots,T$. Only cells in the observed set $\mathcal O$ are available. Series $m\le P$ belong to target markets; the remaining series are helpers. The objective is to estimate missing target cells and report an interval, while leaving observed cells unchanged. The price estimate is obtained by exponentiating the estimated log price; it is not a bias-corrected conditional mean on the price scale.

We distinguish three information regimes. \emph{Retrospective reconstruction} uses observations before and after a gap. \emph{Retrospective filtering} propagates states forward but estimates parameters on the complete supplied panel. \emph{Origin-restricted fitting} refits on the panel prefix ending at the evaluation week, including estimation of weights and interval scales. Only the third can approximate a week-close current-price procedure, and only if preprocessing and helper-release timing respect the same cutoff. We avoid using ``online'' as an unqualified performance claim.

\subsection{Common movement and market-specific persistence}

The working decomposition is
\begin{subequations}\label{eq:model}
\begin{align}
y_{mt} &= a_m+b_m F_{m,t-\ell_m}+e_{mt}, \label{eq:obs}\\
e_{mt} &= \rho_m e_{m,t-1}+\eta_{mt}, \label{eq:residual}\\
\eta_{mt} &\sim \Normal(0,\sigma_{\eta,m}^2).
\end{align}
\end{subequations}
Here $a_m$ is a series-specific level, $b_m$ a factor loading, and $\ell_m$ an optional non-negative lag. The default fixes $b_m=1$ and $\ell_m=0$. The common factor follows a local-level process,
\begin{equation}\label{eq:factor}
f_t=f_{t-1}+u_t,\qquad u_t\sim\Normal(0,q).
\end{equation}
In the pooled candidate, $F_{mt}=f_t$ for every series. A helper-downweighted candidate excludes helpers from the factor's observation updates, although helpers can still influence initialization and parameter updates. This is not equivalent to dropping their columns.

The optional basis candidate instead assigns separate target and helper factor paths:
\begin{subequations}\label{eq:basis}
\begin{align}
f^A_t &= f^N_t+d_t, & d_t&=d_{t-1}+v_t,\\
v_t &\sim\Normal(0,q_d), &
F_{mt}&=\begin{cases}f^A_t,&m\le P,\\f^N_t,&m>P.\end{cases}
\end{align}
\end{subequations}
The helper factor $f^N_t$ follows Equation~\eqref{eq:factor}. The basis variance starts at $q_d=0.1q$ and is subsequently estimated and clipped; the implementation does not enforce $q_d\ll q$ throughout fitting. This candidate permits shared movement without requiring a constant target--helper price difference.

Estimation alternates six rounds of per-series regression and Kalman smoothing. Regression updates the intercept, optional lag, loading, and residual variance; loadings are normalized to mean one and factors are centered to identify scale and level. Factor updates use the fitted residual variance as a working observation-noise variance. Residual persistence is then estimated by a weighted regression of successive observed residual pairs separated by at most four weeks, with weights inversely proportional to their separation, and clipped to $[0,0.97]$. This is a heuristic weekly-persistence estimate, not an exact irregular-gap AR(1) likelihood fit. The overall procedure is an approximate two-stage estimator, not joint maximum likelihood for Equation~\eqref{eq:model}. Implementation settings are collected in Appendix~\ref{app:implementation}.

\subsection{Residual bridges and plug-in uncertainty}

After estimating the factor contribution, define the observed residual as $e_{mt}=y_{mt}-\hat a_m-\hat b_m\hat F_{m,t-\hat\ell_m}$. Consider a missing week bracketed by observed residuals at $s<t<r$. Write $h_L=t-s$, $h_R=r-t$, and $h=h_L+h_R$. For a stationary AR(1) residual with plug-in persistence $\rho_m$, define
\begin{equation}\label{eq:bridgeweights}
\begin{aligned}
A_m &= \frac{\rho_m^{h_L}(1-\rho_m^{2h_R})}{1-\rho_m^{2h}},\\[2pt]
B_m &= \frac{\rho_m^{h_R}(1-\rho_m^{2h_L})}{1-\rho_m^{2h}}.
\end{aligned}
\end{equation}
The conditional residual mean and variance are then
\begin{align}
\hat e_{mt} &= A_m e_{ms}+B_m e_{mr}, \label{eq:bridge}\\
v^e_{mt} &= \tau_m^2\,
\frac{(1-\rho_m^{2h_L})(1-\rho_m^{2h_R})}
     {1-\rho_m^{2h}}, \label{eq:bridgevar}
\end{align}
where $\tau_m^2$ denotes the stationary residual variance, estimated from the observed residuals. These expressions are exact for the stated residual process conditional on a known factor path and parameters. Their use after estimating those quantities does not make the full reconstruction distribution exact.

For a one-sided gap at distance $h$ from its only usable endpoint $s$, the AR candidate uses $\hat e_{mt}=\rho_m^h e_{ms}$ and $v^e_{mt}=\tau_m^2(1-\rho_m^{2h})$. The endpoint must precede $t$ in origin-restricted fitting; a later endpoint is permitted only retrospectively. An alternative random-walk candidate uses
\begin{align}
\hat e^{\mathrm{RW}}_{mt}
&=\frac{h_R}{h}e_{ms}+\frac{h_L}{h}e_{mr},\\
v^{e,\mathrm{RW}}_{mt}&=q_{e,m}\frac{h_Lh_R}{h},
\end{align}
with weekly innovation variance $q_{e,m}$ estimated from squared residual increments divided by their time separation. With only one endpoint it retains that endpoint residual and lets variance grow with distance. Thus the AR candidate allows mean reversion, whereas the random-walk candidate retains a local price difference.

Let $V_{m,t-\ell_m}$ be the fitted factor variance at the relevant lag. The code combines the components as
\begin{align}
\hat y_{mt}&=\hat a_m+\hat b_m\hat F_{m,t-\hat\ell_m}+\hat e_{mt},\\
v_{mt}&=\hat b_m^2 V_{m,t-\hat\ell_m}+v^e_{mt}. \label{eq:pluginvar}
\end{align}
Equation~\eqref{eq:pluginvar} is a plug-in approximation: it omits factor--residual covariance and uncertainty in fitted parameters. Empirical calibration can adjust its scale but does not establish that these omitted terms are negligible.

\subsection{Adaptive ensembling and empirical calibration}\label{sec:auto}

The adaptive method, \auto{}, combines six candidates: the helper-downweighted factor with an AR or random-walk bridge; the basis-linked factor; the pooled factor; and an AR or random-walk bridge without a factor. Basis and pooled-helper candidates are omitted when no helper series are present. The fixed bridge, \fixedbridge{}, uses the helper-downweighted factor, unit loadings, zero lag, and the AR bridge.

For the original ensemble, three inner masks transplant shifted reporting patterns onto target series with at least twenty reports, retaining at least six reports per masked series. These masks are generated after removing the outer evaluation cells. Let $\mathbf y_H$ contain the inner log-price truths and $\mathbf X_H$ the corresponding candidate predictions. With at least sixty inner cells, the implementation computes
\begin{align}
\hat{\mathbf c}
&=\underset{\mathbf c\ge0}{\arg\min}\,
  \|\mathbf y_H-\mathbf X_H\mathbf c\|_2^2,\label{eq:nnls}\\
w_k&=\frac{\hat c_k}{\sum_{j=1}^K\hat c_j}.\label{eq:weights}
\end{align}
If the coefficient sum is zero or too few cells are available, the first factor--AR candidate receives all weight. Importantly, normalization in Equation~\eqref{eq:weights} occurs \emph{after} NNLS; the resulting weights are not claimed to minimize squared error subject to a sum-to-one constraint.

For candidate predictions $\hat y^{(k)}_{mt}$ with variances $v^{(k)}_{mt}$, the ensemble uses
\begin{align}
\bar y_{mt}&=\sum_{k=1}^K w_k\hat y^{(k)}_{mt},\\
\bar v_{mt}&=\sum_{k=1}^K w_k\left[
v^{(k)}_{mt}+(\hat y^{(k)}_{mt}-\bar y_{mt})^2\right].
\end{align}
This working mixture variance combines candidate variance and disagreement; it is not the sampling variance of the weighted estimator.

On inner masked cells, standardized absolute errors are grouped by distance to an observed report: 1, 2--3, 4--8, and at least 9 weeks. With $Q_{0.8}$ denoting an empirical quantile, the variance multiplier in bin $b$ is
\begin{equation}\label{eq:scale}
\lambda_b=\left(\frac{Q_{0.8}(z_i:i\in b)}{1.2816}\right)^2,
\qquad z_i=\frac{|y_i-\bar y_i|}{\sqrt{\bar v_i}}.
\end{equation}
Bins with fewer than thirty cells use the pooled multiplier. The nominal-$100(1-\alpha)\%$ price interval is
\begin{equation}\label{eq:interval}
\left[\exp(\bar y_{mt}-c_\alpha s_{mt}),\;
      \exp(\bar y_{mt}+c_\alpha s_{mt})\right],
\end{equation}
where $s_{mt}^2=\lambda_b\bar v_{mt}$ and $c_\alpha$ is the standard-normal quantile. In particular, $c_{0.2}\simeq1.2816$. Scaling to an inner 80\% target does not calibrate every tail probability. Inner observations also fit the weights, and temporal dependence violates simple exchangeability assumptions; no distribution-free guarantee is asserted.

The revised validation scheme used only in control C expands the inner sample, changes the gap-distance definition for one-sided fills, and modifies scale estimation (Section~\ref{sec:controls}). Because the expanded sample also enters Equation~\eqref{eq:nnls}, it can change point predictions, not only intervals.

\section{Experimental Design}\label{sec:design}

\subsection{Masked reports and development history}

Every scored target is a retained real report that is hidden before fitting. The main scenarios are: a transplant of another series' shifted reporting pattern; blocks of 4, 8, or 13 weeks; the first 26 calendar weeks of a market, retaining at least eight later reports; and state-wide outages of 4 or 8 weeks. Three transplant seeds are used. A separate random-deletion sweep hides 20, 40, 60, or 80\% of observed reports. These experiments assess reconstruction conditional on reports being observed in the source data, not the unobserved prices in the original gaps.

Fixed-bridge settings were selected from twelve configurations using an Assam development run with different masks from the reported evaluation runs. They were then reused in the other states. The adaptive method uses the same candidate family across states but learns its weights within each panel. Different masks do not make the development and evaluation populations independent: both reuse the same underlying series. In addition, the random-walk candidates and adaptive ensemble were designed after earlier results had been inspected. Bihar and Goa therefore illustrate transfer of settings, not a blind external validation.

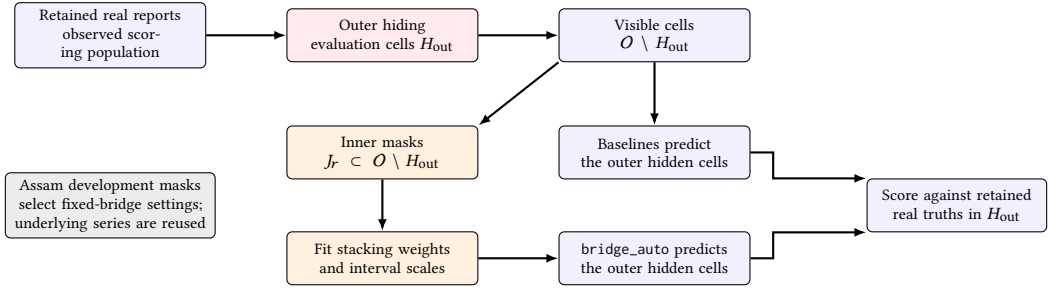
\begin{figure*}[tbp]
\centering
\begin{tikzpicture}[
  box/.style={draw, rounded corners=2pt, align=center, font=\scriptsize, minimum height=7mm, fill=blue!6, inner sep=2.5pt, text width=2.35cm},
  hid/.style={draw, rounded corners=2pt, align=center, font=\scriptsize, minimum height=7mm, fill=red!8, inner sep=2.5pt, text width=2.35cm},
  inn/.style={draw, rounded corners=2pt, align=center, font=\scriptsize, minimum height=7mm, fill=orange!12, inner sep=2.5pt, text width=2.35cm},
  arr/.style={-{Latex[length=1.6mm]}, thick}]
\node[box] (real) at (0,0) {Retained real reports\\observed scoring population};
\node[hid] (outer) at (3.6,0) {Outer hiding\\evaluation cells $H_{\mathrm{out}}$};
\node[box] (vis) at (7.2,0) {Visible cells\\$\mathcal O\setminus H_{\mathrm{out}}$};
\draw[arr] (real) -- (outer);
\draw[arr] (outer) -- (vis);
\node[box] (base) at (7.2,-1.55) {Baselines predict\\the outer hidden cells};
\node[inn] (in1) at (3.6,-1.55) {Inner masks\\$J_r\subset\mathcal O\setminus H_{\mathrm{out}}$};
\node[inn] (w) at (3.6,-2.95) {Fit stacking weights\\and interval scales};
\node[box] (auto) at (7.2,-2.95) {\auto{} predicts\\the outer hidden cells};
\draw[arr] (vis) -- (base);
\draw[arr] (vis.south west) -- (in1.north east);
\draw[arr] (in1) -- (w);
\draw[arr] (w) -- (auto);
\node[box, text width=2.2cm] (score) at (11.2,-2.25) {Score against retained\\real truths in $H_{\mathrm{out}}$};
\draw[arr] (base.east) -- ++(0.3,0) |- (score.north west);
\draw[arr] (auto.east) -- ++(0.3,0) |- (score.south west);
\node[hid, fill=gray!15, text width=2.6cm] at (0,-2.25) {Assam development masks\\select fixed-bridge settings;\\underlying series are reused};
\end{tikzpicture}
\caption{Outer/inner masking schematic. Outer targets are excluded before generating the ensemble's inner masks. Different masks reuse the same source series and do not constitute an untouched test population. The schematic does not certify release-aware preprocessing or eliminate aggregate overlap and result-informed development; the controls address only specified parts of these concerns.}
\Description{Outer hiding produces visible reports used by baselines and the ensemble. Inner masks drawn only from visible reports fit weights and interval scales, while outer truths are reserved for scoring. Assam development is explicitly not an independent population.}
\label{fig:protocol}
\end{figure*}

Figure~\ref{fig:protocol} distinguishes the outer evaluation masks from the inner masks used to fit weights and interval scales. This cell-level separation does not resolve full-sample preprocessing, overlapping aggregates, publication timing, or result-informed model development.

\subsection{Baseline implementations and comparison cohorts}

Ten baselines span simple, statistical, and learned methods. Carry-forward propagates the last report and backfills leading gaps in the retrospective benchmark. Linear interpolation and PCHIP interpolate internally and carry endpoint values. The local-level Kalman baseline selects a process-to-observation variance ratio from six values by likelihood and smooths each series \cite{moritz2017}. SoftImpute uses singular-value shrinkage of $0.15\sigma_1$ for up to 100 iterations \cite{mazumder2010}. The kNN implementation is a study-specific correlation-neighbour heuristic with $k=5$, a minimum of eight common weeks, and median price-offset adjustment; it is not a reproduction of Ma et al.'s forecasting system. Chained ML uses an iterative ExtraTrees imputer with fifty trees and five iterations, representing the broader chained-regression family rather than an exact reproduction of the World Bank model \cite{vanbuuren2011,stekhoven2012,andree2021}.

DFM-EM fits a one-factor dynamic model with AR(1) idiosyncratic components, using at most the sixty best-reported series \cite{banbura2014}. BRITS and SAITS use PyPOTS implementations trained on the supplied panel for fifty epochs, with 24-week windows and stride two. BRITS has a 64-unit recurrent layer; SAITS has two layers, model dimension 64, and four attention heads \cite{cao2018,du2023saits,du2023pypots}. These are fixed-budget implementations, not exhaustive architecture or hyperparameter searches.

Chained ML, DFM-EM, BRITS, and SAITS were run on one instance of each main scenario: transplant seed 0, 8-week block, back-cast, and 8-week outage. Faster methods also ran on the other masks. The main twelve-method table uses the common heavy-method cohort. Pairwise inference uses every run available for its comparator, so its point estimates can differ from the common-cohort table. Within each comparison, scoring is matched by state, commodity, scenario, run, market, and week.

All methods share a numerical guard: a non-finite log-price estimate, or one differing by more than $\log 10$ from the carry-forward/backfilled reference, is replaced by that reference. This is not a nearest-observation rule, and the backward-filled reference is not a past-only estimate. Origin-restricted controls instead use the last earlier report. Fallbacks and unavailable intervals must be distinguished from successful probabilistic predictions.

\subsection{Error measures and dependent observations}

For a matched set $H$ of hidden reports, the principal error measure is
\begin{equation}\label{eq:mape}
\MAPE(H)=\frac{100}{|H|}\sum_{(m,t)\in H}
\frac{|\hat p_{mt}-p_{mt}|}{p_{mt}}.
\end{equation}
All retained prices are positive, but percentage errors remain sensitive to small denominators \cite{hyndman2006}. Median absolute percentage error and mean absolute log error provide complementary summaries. Pooled MAPE weights cells equally, so larger panels contribute more; metric-invariant superiority is not claimed.

The principal sensitivity analysis uses 2,000 bootstrap draws of state--commodity panels within states and common 13-week calendar blocks. Its product resampling follows the pigeonhole-bootstrap rationale \cite{owen2007}, rather than the variance estimator proposed for multiway clustering \cite{cameron2011}. Saved truths are checked against reconstructed source panels before relative week indices are mapped to Monday dates. Blocks share the anchor 5 January 1970, aligning resampled dates across panels. The two-way interval targets the cell-weighted MAPE difference. Paired Wilcoxon tests instead compare panel mean errors, with Holm adjustment within each setting and scope's comparison family \cite{holm1979}. Shared-helper dependence and post-result model development remain unresolved.

Earlier cell-level tests, series-only bootstrap intervals, and Friedman--Nemenyi ranks are retained as descriptive legacy analyses in the appendix \cite{demsar2006}. They are not used as confirmatory evidence: repeated weeks and masks are dependent, series-only resampling omits common market shocks, and the comparison panels are not independent datasets.

\subsection{Component and downstream analyses}

Fixed-bridge ablations remove the regional factor or residual bridge, permit free loadings or lags, alter helper use, or remove interval scaling. These are comparisons of specified implementations. A zero helper observation weight does not isolate the complete removal of helper information; an explicit column-exclusion check is stronger. Ensemble weights are reported separately as descriptive coefficients.

The downstream experiment uses 78 Assam market--commodity series and 1--8-week forecasts from a companion prototype. Histories are completed at each rolling origin by carry-forward or by an ensemble refitted on the corresponding prefix. The forecasters include naive prediction, exponential smoothing, moving average, pooled seasonality, gradient boosting, LightGBM, and Chronos-2. Forecasts are scored only where a real future report exists, using the later half of the evaluated origins. Full-sample cleaning and mid-month placement of monthly helpers remain in that experiment, so it is a retrospective sensitivity analysis rather than a strictly release-aware forecast trial. A hold-or-sell simulation additionally includes storage loss, fees, and interest; simulated returns do not establish a realized benefit to farmers.

\section{Results}\label{sec:results}

\subsection{Retrospective reconstruction and state heterogeneity}

Table~\ref{tab:main} reports the common-cohort comparison. The adaptive ensemble has pooled North-East MAPE of 19.3\%, compared with 26.5\% for SAITS, the lowest-error baseline on that cohort. The difference is approximately 7.2 percentage points, or a 27\% relative reduction. Assam provides a substantial part of the evidence: error is 19.8\% for the ensemble and 30.5\% for SAITS on 27,478 matched hidden cells. The pooled North-East comparison contains approximately 56,000 cells from 1,080 market--commodity series.

\begin{table*}[tbp]
\caption{Retrospective MAPE (\%) on the common heavy-method cohort. Each method is evaluated on the same hidden reports within a state. NE pooled is cell-weighted over the five North-East states. Bold marks the lowest point estimate, not statistical significance.}
\label{tab:main}
\begin{tabular}{lrrrrrrrr}
\toprule
Method & Assam & Meghalaya & Manipur & Nagaland & Tripura & Goa & Bihar & NE pooled \\
\midrule
\auto{} & \textbf{19.8} & \textbf{15.7} & \textbf{13.4} & 19.3 & \textbf{19.6} & \textbf{10.0} & \textbf{14.1} & \textbf{19.3} \\
Fixed bridge & 20.6 & 18.4 & 13.9 & 20.0 & 21.7 & 13.7 & 15.6 & 20.5 \\
SAITS & 30.5 & 17.6 & 16.4 & 20.7 & 24.4 & 13.9 & 22.1 & 26.5 \\
Chained ML & 31.1 & 18.2 & 15.4 & 21.0 & 24.3 & 22.3 & 23.1 & 26.7 \\
BRITS & 32.3 & 18.6 & 17.0 & 21.1 & 27.6 & 15.1 & 24.0 & 28.6 \\
SoftImpute & 38.5 & 20.2 & 20.0 & 24.2 & 31.6 & 22.6 & 29.4 & 33.5 \\
kNN & 38.1 & 31.0 & 20.6 & 32.8 & 29.5 & 13.0 & 21.1 & 33.7 \\
PCHIP & 41.2 & 16.3 & 21.8 & 19.8 & 38.1 & 13.9 & 25.2 & 36.6 \\
Linear & 41.4 & 16.2 & 22.0 & 19.5 & 38.0 & 13.7 & 25.2 & 36.6 \\
Kalman & 41.7 & 16.5 & 22.0 & \textbf{18.7} & 38.2 & 14.2 & 25.4 & 36.7 \\
DFM-EM & 45.0 & 18.6 & 21.8 & 25.2 & 40.6 & 14.1 & 28.1 & 39.8 \\
Carry-forward & 45.9 & 18.6 & 28.0 & 23.1 & 44.8 & 13.8 & 29.5 & 41.8 \\
\bottomrule
\end{tabular}
\end{table*}

The pooled comparison does not imply uniform state-level superiority. The ensemble's common-cohort error is lowest in six states, but its advantage over linear interpolation in Meghalaya is only 0.5 points. In Nagaland, Kalman smoothing is more accurate (18.7\% versus 19.3\%). Different comparison cohorts can also change these orderings: the broader pairwise linear-interpolation cohort in Table~\ref{tab:inference} slightly favours linear interpolation in Meghalaya and Nagaland. Such differences should not be interpreted as contradictions between matched tables or hidden by quoting only the most favourable cohort.

\begin{figure}[tbp]
\centering
\includegraphics[width=\columnwidth]{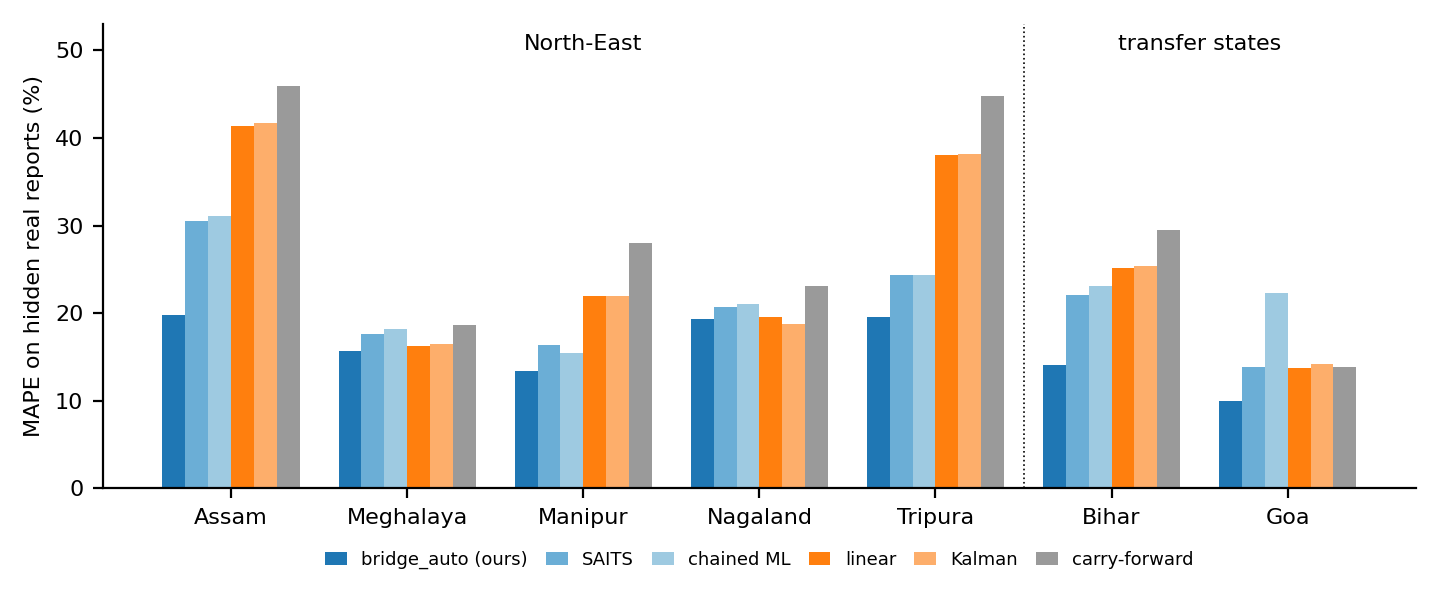}
\caption{Retrospective error by state for selected implementations on the common benchmark. The first five states form the North-East pool; Bihar and Goa are additional comparison states.}
\Description{Grouped bars comparing the ensemble and selected baselines across seven states; gains are uneven and Kalman smoothing is best in Nagaland.}
\label{fig:states}
\end{figure}

The fixed bridge is competitive in Assam, Manipur, Tripura, and Bihar, but performs less well than simple baselines in Meghalaya, Nagaland, and Goa. Adaptive weighting improves its common-cohort point estimates in all seven states. This improvement supports the usefulness of the candidate family on the observed panels, but it is not an independent confirmation of the ensemble design, which was informed by earlier results.

\subsection{Components, borrowing, and gap geometry}

In Assam, removing the regional factor raises fixed-bridge MAPE from 20.6\% to 38.3\%. Removing residual bridging raises it to 22.5\%, permitting free loadings to 24.1\%, and allowing lags to 20.9\% (Table~\ref{tab:ablation}). These contrasts favour a parsimonious shared signal with market-specific persistence in this implementation. They do not identify why an arbitrary factor model would succeed or fail, and initialization prevents treating a zero helper update weight as full helper exclusion.

\begin{figure}[tbp]
\centering
\includegraphics[width=\columnwidth]{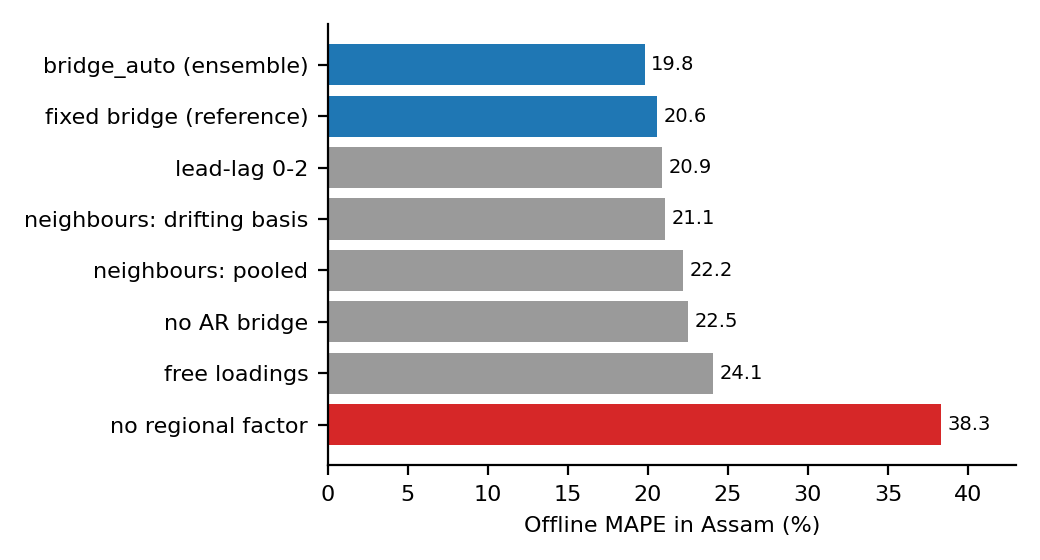}
\caption{Assam fixed-bridge ablations. Each bar represents a specified implementation change; the factor-removal contrast is the largest.}
\Description{Bar chart of Assam reconstruction errors for the fixed bridge and component variants, with the largest increase when the regional factor is removed.}
\label{fig:ablation}
\end{figure}

Estimated ensemble weights differ across states. Basis and pooled-helper candidates receive approximately 66\% of mean weight in Goa, 37--38\% in Manipur and Tripura, and 18--21\% in the other states. The two no-factor candidates receive approximately 45\% in Nagaland and 30\% in Meghalaya. These summaries describe combinations of correlated candidate predictions; they do not prove that a component causes a particular gain, nor that geographic borrowing is uniformly beneficial. Goa's sensitivity to its own monthly average is more informative than its weights alone (Section~\ref{sec:controls}).

Error also depends on the gap geometry. In Assam's main scenarios, adaptive MAPE is 22.9\% for back-casting, 13.2\% for 8-week blocks, 15.4\% for 8-week outages, and 18.8\% for transplant masks. For hidden cells at least nine weeks from an observed report, the fixed bridge has 25.3\% error, compared with 60.3\% for linear interpolation and 40.7\% for SAITS (Table~\ref{tab:gaplen}). With random deletion increasing from 20\% to 80\%, adaptive error rises from 9.5\% to 13.7\%, compared with 10.5\% to 19.2\% for linear interpolation (Table~\ref{tab:missing}). These results concern additional masking of reported values, not measured accuracy in the original unreported weeks.

\begin{figure}[tbp]
\centering
\includegraphics[width=\columnwidth]{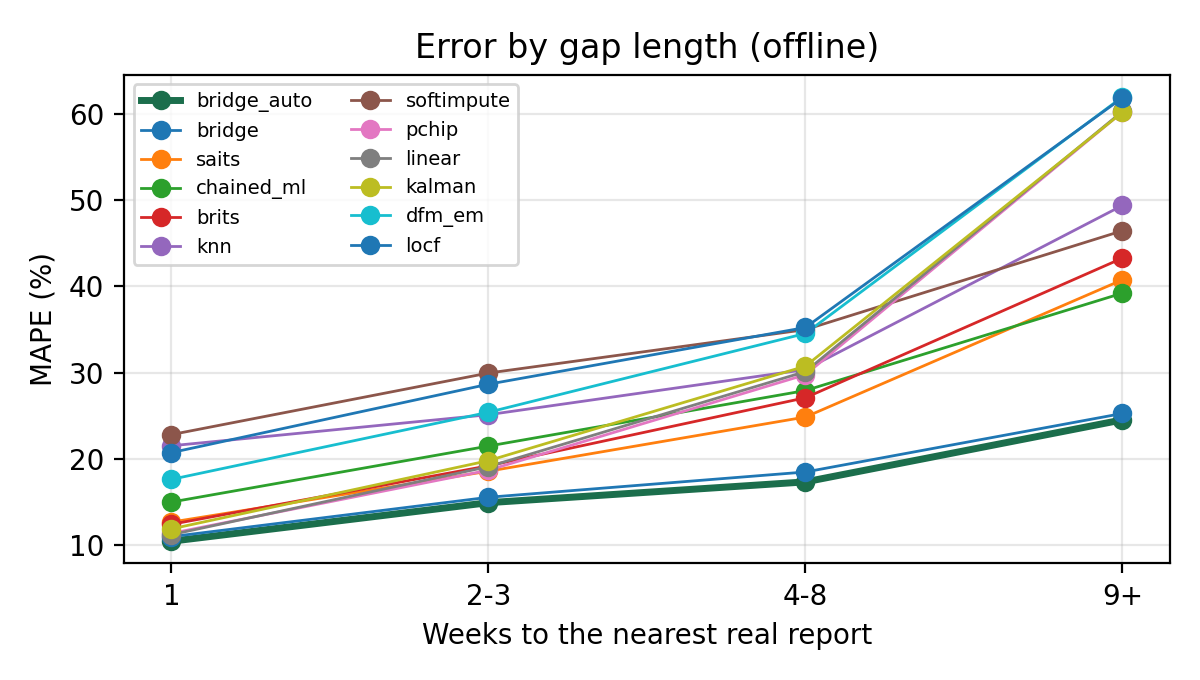}
\caption{Retrospective Assam error by distance from the nearest visible report. The plot assesses deliberately hidden reports, not the unknown original gaps.}
\Description{Error curves against observation distance for the fixed bridge and selected baselines; errors increase as support becomes more distant.}
\label{fig:gaplen}
\end{figure}

The ensemble takes approximately 3.9 seconds per Assam commodity panel on one CPU core, compared with 0.46 seconds for the fixed bridge, 6.0 seconds for SAITS, 18 seconds for BRITS, and 42 seconds for chained ML (Table~\ref{tab:runtime}). These timings characterize the tested configurations and local environment, not general model-family complexity.

\begin{figure*}[tbp]
\centering
\includegraphics[width=0.84\textwidth]{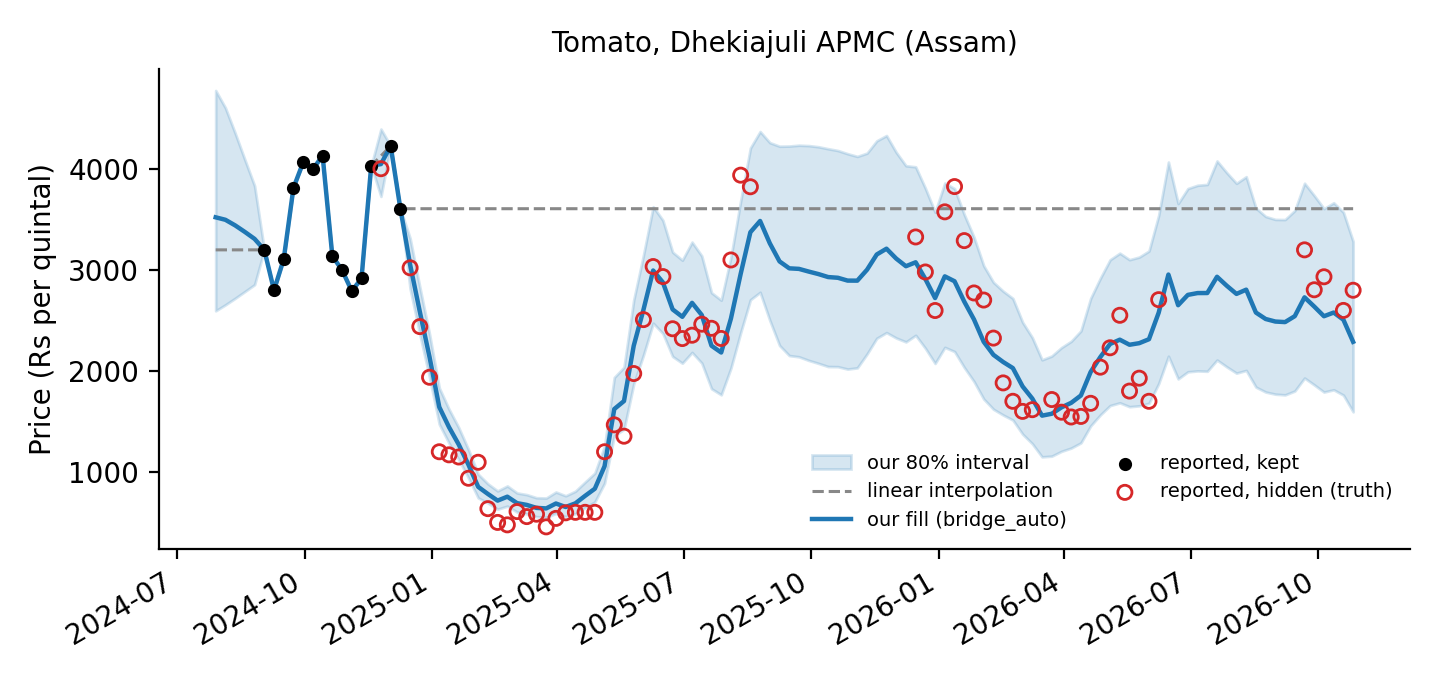}
\caption{Selected long-gap illustration for Tomato at Dhekiajuli APMC, Assam: retained reports, hidden real reports, the adaptive fill and nominal-80\% interval, and linear interpolation. Selection searched Potato, Tomato and Onion over twelve transplant seeds, preferring long gaps with at least fifteen hidden reports and more retained observations. This is an illustrative hard case, not a representative sample or evidence of calibrated uncertainty in original gaps.}
\Description{A selected tomato series shows a prolonged hidden interval where the adaptive estimate changes with the panel signal while linear interpolation remains nearly flat.}
\label{fig:example}
\end{figure*}

Figure~\ref{fig:example} illustrates a selected long-gap reconstruction; it is not a representative accuracy check. Supporting relative-gain, runtime, scenario, and random-deletion plots appear in Appendix~\ref{app:visuals}.

\subsection{Aggregate exclusion and information timing}\label{sec:controls}

Three controls separate distinct sources of sensitivity. They differ in information processing and, for C, in the ensemble-validation procedure. Their raw pooled estimates must not be compared as if all used the same scoring population.

\paragraph{A: removing the target-state monthly aggregate.}
All twelve methods are refitted on the common heavy-method runs without the target state's monthly average, retaining the original target cells. Pooled North-East adaptive MAPE changes from 19.3\% to 19.4\% (Table~\ref{tab:controls}). This supports a limited conclusion: the main pooled advantage does not depend strongly on that aggregate in the observed benchmark. It does not establish absence of all information leakage. In Goa, adaptive error increases from 10.0\% to 13.9\%, similar to linear interpolation at 13.7\%. The original Goa advantage therefore depends materially on the retained state aggregate and should not be treated as robust transfer evidence.

\paragraph{B: origin-restricted fitting with retrospective preprocessing.}
For transplant seed 0 and the 8-week outage on up to ten commodities per state, the ensemble, fixed bridge, Kalman baseline, and forward-only carry-forward are evaluated at each week using a panel prefix. Monthly helpers enter in the week containing the seventh day after month end; the target's own aggregate is excluded. This assumes publication within seven days and is not verified against actual releases. Cleaning and market/commodity eligibility still use the complete sample. On 21,417 North-East cells, adaptive MAPE is 19.0\%, compared with 37.1\% for carry-forward and 37.7\% for Kalman filtering. The fixed bridge is marginally better pooled (18.97\% versus 19.03\%), showing that adaptive weighting is not uniformly helpful with short histories. Learned baselines were not refitted at each origin.

\begin{table*}[t]
\caption{Aggregate-exclusion and timing controls (MAPE, \%). A removes the target state's monthly average on the matched heavy-method cohort. B refits the four displayed methods at each origin on up to ten commodities per state, excluding the target aggregate and assigning monthly helpers an assumed release week. B retains retrospective cleaning and eligibility. Bold marks the lowest point estimate within each block; Cov. is empirical nominal-80\% coverage for \auto{}. A and B have different scoring cohorts.}
\label{tab:controls}
\small
\setlength{\tabcolsep}{4.5pt}
\begin{tabular}{l rrl rrrr rr}
\toprule
 & \multicolumn{3}{c}{A. no own monthly average (offline)} & \multicolumn{6}{c}{B. origin-restricted fitting} \\
\cmidrule(lr){2-4}\cmidrule(lr){5-10}
State & \auto{} with & \auto{} without & best baseline & \auto{} & bridge & carry-fwd & Kalman & Cov. & Cells \\
\midrule
Assam & 19.8 & \textbf{19.8} & 30.1 (SAITS) & \textbf{18.4} & 18.6 & 46.5 & 48.1 & 79.2 & 8473 \\
Meghalaya & 15.7 & \textbf{15.9} & 16.2 (linear) & 17.7 & 17.8 & \textbf{17.0} & 17.7 & 58.1 & 1381 \\
Manipur & 13.4 & \textbf{13.8} & 17.0 (SAITS) & 15.3 & \textbf{14.8} & 30.3 & 31.0 & 59.7 & 918 \\
Nagaland & 19.3 & 19.4 & 18.7 (Kalman) & 23.2 & \textbf{22.8} & 24.0 & 23.3 & 60.9 & 2418 \\
Tripura & 19.6 & \textbf{20.0} & 24.3 (chained ML) & 19.1 & \textbf{18.9} & 35.4 & 35.4 & 74.9 & 8227 \\
NE pooled & 19.3 & \textbf{19.4} & 26.5 (SAITS) & 19.0 & \textbf{19.0} & 37.1 & 37.7 & 73.3 & 21417 \\
\midrule
Bihar & 14.1 & \textbf{15.4} & 21.1 (kNN) & \textbf{16.7} & 18.1 & 32.2 & 32.1 & 62.8 & 924 \\
Goa & 10.0 & 13.9 & 13.7 (linear) & \textbf{8.6} & 9.2 & 9.1 & 10.9 & 90.0 & 40 \\
\bottomrule
\end{tabular}
\end{table*}

\paragraph{C: week-close preprocessing and revised inner validation.}
Hidden market--week reports are removed from the raw daily inputs before cleaning. The cleaner uses a trailing 43-day market window, an expanding state--commodity median, and the same-week cross-market median available at week close. Eligibility thresholds are enforced on visible reports through each origin. This is a week-close procedure, not a day-by-day nowcast: later days in the current week may contribute to its cross-market median. The scoring truths and commodity shortlist still derive from the retrospective pipeline.

Control C also expands inner masking to sparser target and helper series when necessary, calibrates from twenty inner cells, measures one-sided gap distance from the previous report, and shrinks bin-specific variance scales toward a pooled scale. It uses a finite-sample-adjusted quantile level implemented with interpolated empirical quantiles, not a split-conformal guarantee. Because the expanded masks also fit stacking weights, C is a combined input-processing and ensemble-validation intervention.

On C's 14,103 North-East cells, adaptive MAPE is 19.7\%, compared with 30.8\% for carry-forward and 31.1\% for Kalman filtering (Table~\ref{tab:causal}). Simple methods are better in Meghalaya and Nagaland. The smaller cohort reflects the origin-wise eligibility rules. Comparing C's 19.7\% directly with B's unmatched 19.0\% would mix changes in inputs, ensemble fitting, and scoring population.

Table~\ref{tab:matchedprep} instead compares saved B and C predictions on identical C cells. With fixed-bridge model settings unchanged, MAPE rises from 18.62\% to 20.46\%; this isolates the input/preprocessing change for that fixed implementation. The adaptive ensemble changes from 18.68\% to 19.75\%, but that contrast includes revised weight fitting. It cannot identify the adaptive model's isolated response to preprocessing or calibration.

\begin{table*}[tbp]
\caption{Combined pipeline C (MAPE, \%): origin-restricted fitting, week-close input processing, origin-wise eligibility, and revised inner validation. Hidden reports are removed before cleaning; within-week observations remain available at week close. Bold marks the lowest point estimate. The right-hand columns compare the adaptive ensemble under B and C on the same C cells. Coverage changes therefore reflect a combined intervention, not an isolated calibration effect. Cov80 and Cov95 are empirical coverage at the corresponding nominal levels.}
\label{tab:causal}
\small
\setlength{\tabcolsep}{5pt}
\begin{tabular}{l rrrr r r rr rr}
\toprule
 & \multicolumn{5}{c}{C: combined pipeline} & \multicolumn{5}{c}{\auto{} on C's cells: B against C} \\
\cmidrule(lr){2-6}\cmidrule(lr){7-11}
State & \auto{} & bridge & carry-fwd & Kalman & Cells & MAPE B & Cov80 B & Cov80 C & Cov95 B & Cov95 C \\
\midrule
Assam & \textbf{22.7} & 23.3 & 38.0 & 39.0 & 5302 & 18.5 & 83.6 & 81.7 & 93.0 & 93.8 \\
Meghalaya & 18.4 & 16.2 & 14.3 & \textbf{14.2} & 578 & 16.0 & 60.7 & 80.1 & 72.5 & 91.7 \\
Manipur & 12.8 & \textbf{12.2} & 18.5 & 19.1 & 736 & 12.6 & 62.2 & 70.1 & 72.7 & 80.6 \\
Nagaland & 21.9 & 21.6 & 21.1 & \textbf{20.1} & 1376 & 21.8 & 65.1 & 73.9 & 81.8 & 87.6 \\
Tripura & \textbf{17.7} & 19.2 & 29.8 & 29.8 & 6111 & 19.1 & 76.4 & 77.8 & 88.1 & 90.3 \\
NE pooled & \textbf{19.7} & 20.5 & 30.8 & 31.1 & 14103 & 18.7 & 76.6 & 78.6 & 87.9 & 90.9 \\
\midrule
Bihar & \textbf{16.6} & 19.5 & 28.9 & 29.1 & 605 & 16.8 & 66.3 & 70.7 & 79.0 & 84.8 \\
Goa & \textbf{9.7} & 10.4 & 10.4 & 12.6 & 32 & 9.7 & 93.8 & 93.8 & 100.0 & 100.0 \\
\bottomrule
\end{tabular}
\end{table*}

\begin{table*}[t]
\caption{Matched B--C MAPE (\%) on C's cells, using saved predictions. Fixed bridge uses identical model settings in both arms, isolating the input/preprocessing intervention for that model. Adaptive ensemble C also changes inner validation and stacking weights, so its contrast is a combined intervention. Change is C minus B, in percentage points.}
\label{tab:matchedprep}
\small
\begin{tabular}{lrrrrrrr}
\toprule
 & \multicolumn{3}{c}{Fixed bridge: input change} & \multicolumn{3}{c}{Adaptive ensemble: combined change} & \\
State & B & C & Change & B & C & Change & Cells \\
\midrule
Assam & 18.86 & 23.27 & 4.40 & 18.51 & 22.71 & 4.20 & 5302 \\
Meghalaya & 16.05 & 16.22 & 0.17 & 15.96 & 18.36 & 2.40 & 578 \\
Manipur & 11.97 & 12.17 & 0.20 & 12.55 & 12.80 & 0.25 & 736 \\
Nagaland & 20.80 & 21.65 & 0.84 & 21.81 & 21.88 & 0.06 & 1376 \\
Tripura & 18.97 & 19.16 & 0.19 & 19.12 & 17.67 & -1.46 & 6111 \\
Bihar & 19.15 & 19.51 & 0.36 & 16.77 & 16.63 & -0.14 & 605 \\
Goa & 10.36 & 10.36 & 0.00 & 9.68 & 9.68 & -0.00 & 32 \\
North-East pooled & 18.62 & 20.46 & 1.83 & 18.68 & 19.75 & 1.06 & 14103 \\
\bottomrule
\end{tabular}
\end{table*}

\subsection{Dependence-aware sensitivity intervals}\label{sec:inference}

The corrected panel/time-block analysis supports a positive pooled retrospective difference against SAITS: 7.2 percentage points, with a 95\% interval of 2.8--12.7 (Table~\ref{tab:inference}). The interval against linear interpolation is 4.4--14.5 points, around a 9.1-point difference on its broader pairwise cohort. All ten pooled offline comparison intervals exclude zero, but those rows do not share one universal scoring cohort. For each state's closest comparator in the pairwise analysis, only Assam and Bihar have two-way intervals entirely above zero.

The corresponding pooled carry-forward differences are 18.1 points (8.5--28.6) in B and 11.0 points (5.8--17.0) in C. These intervals are wider than series-only resampling intervals because they address additional dependence. They remain exploratory: relatively few calendar blocks and panels are available, helpers induce links across states, and the design was not fixed before all results were inspected. Positive intervals should not be described as independent confirmation or evidence of a significant advantage in every state.

\begin{table*}[t]
\caption{Exploratory dependence-aware comparisons on matched cells. Diff. is comparator minus adaptive MAPE (points). Commodity CI resamples state--commodity panels within states; Two-way CI also resamples common 13-week calendar blocks anchored on 5 January 1970 (95\%, 2,000 draws). Panels counts positive panel-mean differences. Wilcoxon $p$ values are Holm-adjusted across comparisons within each setting and scope, before selecting the closest state comparator. Bootstrap and panel-test estimands differ. Offline cohorts vary by comparator and need not match Table~\ref{tab:main}. C includes both input-processing and ensemble-validation changes.}
\label{tab:inference}
\small
\setlength{\tabcolsep}{4.5pt}
\begin{tabular}{lll rrr cc rr}
\toprule
Setting & Scope & Comparison & \auto{} & Other & Diff. & Commodity CI & Two-way CI & Panels & $p$ \\
\midrule
offline & NE pooled & SAITS & 19.3 & 26.5 & 7.2 & 4.7 to 10.0 & 2.8 to 12.7 & 57/68 & $<$0.001 \\
 &  & chained ML & 19.3 & 26.7 & 7.5 & 5.1 to 10.1 & 2.8 to 13.6 & 56/68 & $<$0.001 \\
 &  & linear & 14.7 & 23.8 & 9.1 & 6.1 to 12.4 & 4.4 to 14.5 & 47/68 & $<$0.001 \\
 &  & PCHIP & 14.7 & 23.8 & 9.1 & 6.2 to 12.4 & 4.6 to 14.3 & 47/68 & $<$0.001 \\
 &  & BRITS & 19.3 & 28.6 & 9.3 & 6.5 to 12.5 & 4.4 to 16.3 & 60/68 & $<$0.001 \\
 &  & Kalman & 14.7 & 24.8 & 10.1 & 6.9 to 13.8 & 5.0 to 16.1 & 49/68 & $<$0.001 \\
 &  & kNN & 14.7 & 27.1 & 12.4 & 10.8 to 14.2 & 10.4 to 14.5 & 68/68 & $<$0.001 \\
 &  & SoftImpute & 14.7 & 27.1 & 12.4 & 10.0 to 15.2 & 8.2 to 18.4 & 68/68 & $<$0.001 \\
 &  & carry-forward & 14.7 & 30.1 & 15.4 & 11.7 to 19.4 & 8.9 to 22.4 & 64/68 & $<$0.001 \\
 &  & DFM-EM & 19.3 & 39.8 & 20.5 & 14.2 to 28.1 & 10.9 to 31.8 & 61/68 & $<$0.001 \\
\midrule
origin-restr. (B) & NE pooled & fixed bridge & 19.0 & 19.0 & $-0.1$ & $-0.3$ to 0.2 & $-0.5$ to 0.3 & 23/48 & 0.931 \\
 &  & carry-forward & 19.0 & 37.1 & 18.1 & 12.5 to 23.0 & 8.5 to 28.6 & 30/48 & $<$0.001 \\
 &  & Kalman & 19.0 & 37.7 & 18.7 & 12.8 to 23.7 & 8.7 to 29.8 & 33/48 & $<$0.001 \\
\midrule
+ combined (C) & NE pooled & fixed bridge & 19.7 & 20.5 & 0.7 & $-0.3$ to 2.2 & $-0.6$ to 3.4 & 25/48 & 0.899 \\
 &  & carry-forward & 19.7 & 30.8 & 11.0 & 7.2 to 15.3 & 5.8 to 17.0 & 29/48 & 0.039 \\
 &  & Kalman & 19.7 & 31.1 & 11.4 & 7.5 to 15.7 & 6.0 to 17.6 & 29/48 & 0.039 \\
\midrule
offline, closest & Assam & SAITS & 19.8 & 30.5 & 10.7 & 7.4 to 14.3 & 4.8 to 17.0 & 19/20 & $<$0.001 \\
 & Meghalaya & linear & 13.2 & 12.7 & $-0.4$ & $-1.4$ to 0.3 & $-1.6$ to 0.7 & 5/12 & 0.904 \\
 & Manipur & chained ML & 13.4 & 15.4 & 2.0 & $-0.1$ to 4.2 & $-0.7$ to 5.4 & 7/8 & 0.234 \\
 & Nagaland & linear & 15.9 & 15.4 & $-0.5$ & $-1.2$ to 0.1 & $-1.6$ to 0.3 & 5/12 & 0.549 \\
 & Tripura & chained ML & 19.6 & 24.3 & 4.7 & 1.3 to 9.3 & $-0.9$ to 11.9 & 13/16 & 0.004 \\
 & Bihar & kNN & 10.9 & 17.2 & 6.3 & 2.9 to 9.1 & 2.3 to 9.8 & 11/11 & 0.010 \\
 & Goa & linear & 9.4 & 10.1 & 0.7 & 0.1 to 1.4 & $-0.4$ to 2.2 & 4/5 & -- \\
\bottomrule
\end{tabular}
\end{table*}

\subsection{Interval reliability and model-designated coverage}

In Assam's offline benchmark, nominal 50\%, 80\%, and 95\% intervals contain 54.8\%, 79.5\%, and 90.7\% of hidden reports, respectively. Calibration near the 80\% target does not repair the upper nominal level. For observations at least nine weeks from a report, nominal-80\% coverage falls to approximately 65--67\% in Meghalaya and Manipur; it is 69\% at distances of 4--8 weeks in Bihar (Table~\ref{tab:calib}).

\begin{figure}[tbp]
\centering
\includegraphics[width=\columnwidth]{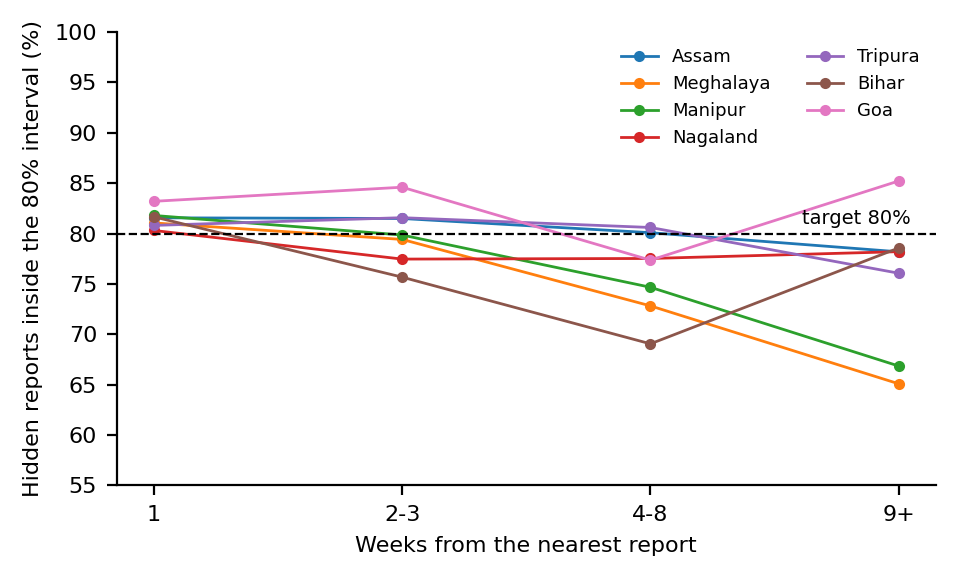}
\caption{Empirical coverage of nominal-80\% retrospective intervals by state and distance to an observed report. Coverage below the target indicates underestimated uncertainty.}
\Description{Coverage curves for seven states and four observation-distance bins, showing undercoverage in several small panels and long gaps.}
\label{fig:calibration}
\end{figure}

Under C, pooled nominal-80\% coverage is 78.6\% and nominal-95\% coverage is 90.9\%. State-level nominal-80\% coverage remains only 70.1\% in Manipur, 73.9\% in Nagaland, and 70.7\% in Bihar. Although some observed coverage improves relative to B on matched cells, this is a combined-pipeline contrast, not an identified calibration-only effect. Goa's 32 scored cells are insufficient to support a broad reliability claim.

Table~\ref{tab:coverage} reports a separate width-based diagnostic: a cell is counted when it has a report or an estimate whose nominal-80\% interval lies within $\pm15\%$. It uses broader eligibility than Table~\ref{tab:datastats}. The resulting share rises from 22.5\% to 42.1\% in Meghalaya and from 28.2\% to 31.5\% in Assam. These are \emph{model-designated} coverage increases, not verified additions to trustworthy price observations. Given the undercoverage, interval width alone cannot certify an imputed price for advice or trading.

\begin{table}[tbp]
\caption{Width-based, model-designated coverage (\%). A cell contains a report or an estimate meeting the nominal-80\%, $\pm15\%$ interval-width rule. Markets need only four reports, and all commodities with 150 daily rows are admitted; this population differs from Table~\ref{tab:datastats}.}
\label{tab:coverage}
\small
\begin{tabular}{lrr}
\toprule
State & Reported & Reported or width-qualified \\
\midrule
Assam & 28.2 & 31.5 \\
Meghalaya & 22.5 & 42.1 \\
Manipur & 75.9 & 83.3 \\
Nagaland & 30.7 & 42.4 \\
Tripura & 59.0 & 65.5 \\
Goa & 98.6 & 98.6 \\
Bihar & 16.0 & 30.5 \\
\bottomrule
\end{tabular}
\end{table}

\subsection{Downstream sensitivity}

Completing histories with the origin-refitted ensemble modestly improves all seven individual forecast point estimates in the saved Assam experiment (Table~\ref{tab:downstream}). The prototype's model-selection engine changes from 18.37\% to 17.57\% MAPE on 21,725 paired forecast cells. Its 0.80-point difference has a saved series-cluster interval of 0.36--1.27 points, which does not address overlapping forecast horizons or common calendar shocks. The naive forecaster remains more accurate overall than the model-selection engine: 16.64\% with completed histories versus 17.57\% for the engine. These results therefore concern input-history sensitivity, not superiority of the forecasting engine.

\begin{table}[tbp]
\caption{Saved Assam forecast sensitivity results: MAPE (\%) on paired later rolling origins, with carry-forward or ensemble-completed input histories. Changes use unrounded saved values. Input cleaning and helper-release timing are not fully origin-restricted.}
\label{tab:downstream}
\small
\setlength{\tabcolsep}{4pt}
\begin{tabular}{lrrr}
\toprule
Forecaster & Carry-forward & Completed & Change \\
\midrule
Naive & 16.88 & 16.64 & $-0.25$ \\
Exponential smoothing & 17.78 & 17.43 & $-0.35$ \\
Chronos-2 & 18.16 & 17.24 & $-0.92$ \\
LightGBM (global) & 18.04 & 17.36 & $-0.68$ \\
Gradient boosting (global) & 18.50 & 17.86 & $-0.65$ \\
Pooled seasonality & 19.96 & 19.78 & $-0.18$ \\
Moving average & 21.25 & 20.46 & $-0.79$ \\
Selection engine & 18.37 & 17.57 & $-0.80$ \\
\bottomrule
\end{tabular}
\end{table}

The hold-or-sell simulation contains 817 decisions. Mean gain relative to immediate sale changes from $-0.21\%$ with carry-forward histories to $+0.29\%$ with ensemble-completed histories, after the simulated costs. However, the share of decisions worse than immediate sale rises from 3.3\% to 7.7\%. The small mean improvement and increased downside frequency do not establish a reliable decision benefit. Full-sample cleaning, assumed helper timing, and the simulation's economic assumptions further limit interpretation.

\section{Discussion}\label{sec:discussion}

The strongest result is a pooled retrospective advantage for a parsimonious factor-and-residual representation under the tested configurations. A common price movement can provide information during gaps when a market's own series offers little local support. The Assam ablation is consistent with this explanation, but it does not uniquely establish the mechanism. Training budget, commodity composition, helper selection, and the pattern of observed reports all remain possible contributors to the relative results.

State heterogeneity is consequential rather than incidental. Meghalaya and Nagaland are sparse, yet simple methods remain competitive or better. Sparsity alone therefore does not identify where the ensemble should be preferred. A practical selection rule would need prospective evidence linking reporting geometry and market co-movement to comparative performance. Until that evidence exists, the current results support investigation of retrospective history completion, not an automatic regional current-price service.

\section{Limitations}\label{sec:limits}

\paragraph{Unobserved prices and measurement error.}
Evaluation targets reports that actually exist, after cleaning. It does not observe the prices in the original gaps or establish that trade occurred there. Informative reporting, market closures, residual entry errors, and unverified unit consistency can all change the target population. The available history spans roughly two years, providing limited seasonal replication.

\paragraph{Adaptation and transfer.}
Development and evaluation masks reuse underlying series, and the ensemble was modified after earlier results were examined. Other-state results reuse this design history, so they are not blind transfer tests. Geography-based helpers are not verified trade partners. A future confirmatory study should freeze the method and compare it on a later period or genuinely untouched markets or commodities.

\paragraph{Information availability and control scope.}
Retrospective smoothing is not current-price estimation. B restricts fitting but retains full-sample cleaning and eligibility. C improves week-close input processing but retains retrospectively defined scoring truth and commodity selection, and changes inner validation at the same time. Monthly publication delays are assumed, not measured. Learned baselines are absent from the origin-wise controls, and C's isolated adaptive preprocessing effect has not been estimated. These limitations cannot be repaired by changing the terminology alone.

\paragraph{Uncertainty and inference.}
The plug-in variance omits covariance and parameter-estimation terms; the same inner cells fit weights and interval scales. Outer coverage is inadequate in several states and at longer distances. The calendar bootstrap addresses one important dependence structure but not all shared-helper links or model-selection uncertainty. Panel tests and cell-weighted bootstrap differences have distinct estimands, and neither is a guarantee of reproducible prospective superiority.

\paragraph{Scale and reproducibility.}
The benchmark selects at most twenty commodities per state, uses three transplant seeds, and evaluates expensive baselines on a subset of runs. Results depend on fixed training budgets and fallback rules. The released study code includes self-contained daily/monthly loaders, tests, pinned environment records, and a saved-result verification command. Complete public empirical replication still requires lawful access to the matching price snapshots or prediction artifacts. An immutable archival DOI, independently verified retrieval times, and a self-contained downstream forecasting engine remain unavailable; successful local tests do not establish prospective validity.

\section{Conclusion}

A regional factor combined with market-specific residual bridging offers a useful reconstruction strategy for the selected short, sparse mandi panels. The adaptive ensemble reduces common-cohort pooled North-East MAPE from 26.5\% for SAITS to 19.3\%, with a positive calendar-block sensitivity interval for the difference. Aggregate-exclusion and origin-restricted controls qualify rather than erase that result: the pooled advantage persists, but Goa's original gain depends on its state aggregate, and simple methods remain preferable in some states. Interval undercoverage and the absence of untouched confirmation prevent a deployment claim. The next scientific step is a frozen, release-aware evaluation with origin-wise learned baselines and independently assessed uncertainty, not a stronger interpretation of the existing retrospective comparisons.

\section*{Data and Code Availability}
Study code, tests, environment records, and reproduction instructions are hosted at \url{https://github.com/khyontekai-ai/price-gap-study}. The original software is MIT-licensed; this licence does not apply to the source price data or manuscript. The release includes daily/monthly loaders without a neighbouring-checkout dependency, source-snapshot checksums, and historical request templates. Daily responses were obtained from \url{https://api.agmarknet.gov.in/v1/dashboard-data/}; monthly responses from \url{https://api.agmarknet.gov.in/v1/price-trend/wholesale-prices-monthly}. Raw snapshots and row-level predictions are not publicly redistributed while source-data permissions remain unverified. The no-training reproduction path requires those matching artifacts; code availability alone therefore does not provide complete public empirical replication. No archival DOI or complete timestamped HTTP-response history is claimed.

\section*{AI-Assisted Preparation}
AI-assisted tools were used for code review, bibliographic checks, and revision of prose and LaTeX. The author remains responsible for the claims, references, numerical results, and final manuscript. This assistance does not constitute independent scientific validation.

\FloatBarrier
\onecolumn
\begin{multicols}{2}
\begingroup
\interlinepenalty=10000
\bibliographystyle{ACM-Reference-Format}
\bibliography{refs_audited}
\endgroup
\end{multicols}
\appendix
\begingroup
\newenvironment{appendixflowtable}{%
  \par\addvspace{7pt}\noindent\begin{minipage}{\linewidth}%
  \captionsetup{type=table}\centering
}{\end{minipage}\par\addvspace{5pt}}
\renewenvironment{table}[1][]{\begin{appendixflowtable}}{\end{appendixflowtable}}
\renewenvironment{table*}[1][]{\begin{appendixflowtable}}{\end{appendixflowtable}}
\section{Implementation Details}\label{app:implementation}

For each panel, the implementation initializes intercepts from observed means and the factor from the centered cross-sectional mean, interpolating missing initialization weeks. Six alternating rounds update regressions, factor states, and process variances. Residual variance has a numerical floor of $10^{-6}$; regression observation variance and factor innovation variance have floors of $10^{-5}$. Free loadings are clipped before normalization. These implementation choices are held fixed in the reported comparison, but the full ensemble design was not specified before all exploratory results.

\begin{table}[t]
\caption{Settings of the implemented retrospective pipeline. Fixed-bridge selection used the Assam development masks; ensemble design also used earlier exploratory results. Control C changes eligibility, input processing, and inner validation as described in the text.}
\label{tab:settings}
\small
\begin{tabular}{p{0.18\linewidth}p{0.75\linewidth}}
\toprule
Part & Setting \\
\midrule
Panel & Target market $\ge10$ reported weeks; helper market $\ge20$; monthly state series $\ge3$ months; commodity $\ge150$ reported days in the state, at most 20 per state \\
Cleaning & Factor 3 against own median ($\pm21$ days) and state-week median; factor 8 against long-run state median \\
Factor model & 6 rounds; loadings clipped to $[0.2,3]$ (or fixed at 1); lags $\{0\}$ or $\{0,1,2\}$; at least 6 reports per series; factor variance floor $10^{-5}$; basis drift variance $0.1q$ at start \\
AR bridge & $\rho_m\in[0,0.97]$ from residual pairs at gaps $\le4$ weeks, weighted $1/g$ \\
Inner validation & 3 transplant masks; masked markets with $\ge20$ reports; donors reporting 15--90\% of weeks; $\ge6$ reports kept; at least 60 inner cells, else candidate 1 only \\
Stacking & NNLS on the hidden truth, weights rescaled to sum to one \\
Calibration & Gap bins 1, 2--3, 4--8, 9+ weeks; scale $(Q_{0.8}(z)/1.2816)^2$; bins with $<30$ cells use the pooled value \\
Safety rule & Replace non-finite fills or a log-price departure greater than $\log 10$ from the carry-forward/backfilled reference; origin-wise controls use the last earlier report \\
\bottomrule
\end{tabular}
\end{table}

\section{Supplementary Comparison Tables}
The following tables retain the saved experiments, with their original numerical results. ``Online'' in a legacy column means retrospective filtering with full-panel parameter fitting, not a past-only experiment. The legacy cell-level tests and series-only intervals are descriptive and should not supersede Table~\ref{tab:inference}. Ablation and distance tables can use broader run cohorts than the main heavy-method comparison.

\begin{table*}[tbp]
\caption{Retrospective filtered-state MAPE (\%) on hidden reports. Parameters and weights use the supplied full panel; these are not strict past-only results.}
\label{tab:online}
\small
\begin{tabular}{lrrrrrrrr}
\toprule
Method & Assam & Meghalaya & Manipur & Nagaland & Tripura & Goa & Bihar & NE pooled \\
\midrule
\auto{} & \textbf{21.3} & \textbf{18.0} & \textbf{10.6} & \textbf{20.8} & \textbf{18.4} & \textbf{11.2} & \textbf{14.6} & \textbf{19.7} \\
bridge (fixed) & 21.7 & 20.4 & 11.8 & 22.2 & 19.4 & 17.3 & 15.4 & 20.5 \\
carry-forward & 44.7 & 18.2 & 23.1 & 22.6 & 41.2 & 13.3 & 29.1 & 39.4 \\
Kalman & 44.7 & 18.8 & 22.9 & 21.6 & 41.7 & 13.0 & 28.8 & 39.5 \\
DFM-EM & 44.7 & 18.7 & 23.0 & 25.9 & 41.3 & 14.4 & 28.3 & 40.0 \\
\bottomrule
\end{tabular}
\end{table*}

\begin{table*}[tbp]
\caption{Exploratory paired comparisons, North-East pooled, offline ($\approx$56{,}000 matched hidden cells, 1{,}080 series). Cell-level Holm-adjusted Wilcoxon $p$ values assume independence not met here; series-bootstrap intervals omit cross-market dependence.}
\label{tab:tests}
\small
\begin{tabular}{lrrrl}
\toprule
Baseline & Its MAPE & Ours better (\% of cells) & Holm $p$ & MAPE gain, points (95\% CI) \\
\midrule
SAITS & 26.5 & 58.6 & $<0.001$ & 7.2 (6.2 to 8.4) \\
chained ML & 26.7 & 58.1 & $<0.001$ & 7.5 (6.5 to 8.6) \\
BRITS & 28.6 & 61.3 & $<0.001$ & 9.3 (8.2 to 10.6) \\
SoftImpute & 33.5 & 68.5 & $<0.001$ & 14.2 (12.9 to 15.5) \\
kNN & 33.7 & 65.9 & $<0.001$ & 14.5 (13.2 to 15.9) \\
linear & 36.6 & 63.7 & $<0.001$ & 17.3 (15.5 to 19.0) \\
PCHIP & 36.6 & 63.4 & $<0.001$ & 17.3 (15.5 to 19.0) \\
Kalman & 36.7 & 64.8 & $<0.001$ & 17.5 (15.7 to 19.3) \\
DFM-EM & 39.8 & 64.7 & $<0.001$ & 20.5 (18.6 to 22.6) \\
carry-forward & 41.8 & 66.8 & $<0.001$ & 22.5 (20.5 to 24.5) \\
\bottomrule
\end{tabular}
\end{table*}

\begin{table*}[tbp]
\caption{Ablations in Assam: offline MAPE (\%) by scenario, offline median APE, and retrospective filtered-state MAPE. Variants change one part of the fixed bridge. Dropping the neighbour markets, switching neighbours off or removing calibration leaves the point fill of the fixed bridge unchanged.}
\label{tab:ablation}
\small
\begin{tabular}{lrrrrrrr}
\toprule
Variant & Back-cast & Block & Outage & Transplant & All & Median APE & Online all \\
\midrule
\auto{} & \textbf{22.9} & \textbf{13.2} & \textbf{15.4} & \textbf{18.8} & \textbf{19.8} & \textbf{14.5} & \textbf{21.3} \\
bridge (fixed) & 23.6 & 13.9 & 15.7 & 19.8 & 20.6 & 15.2 & 21.7 \\
lead-lag 0--2 & 23.9 & 14.2 & 15.6 & 20.2 & 20.9 & 15.3 & 23.2 \\
neighbours: drifting basis & 23.4 & 14.2 & 16.2 & 20.9 & 21.1 & 15.5 & 21.8 \\
neighbours: pooled & 24.4 & 15.3 & 17.1 & 22.3 & 22.2 & 16.3 & 23.3 \\
target-state markets only & 23.1 & 13.7 & 15.6 & 20.5 & 20.7 & 15.1 & 22.0 \\
no AR bridge & 24.7 & 18.1 & 19.7 & 21.7 & 22.5 & 16.9 & 22.7 \\
free loadings & 27.0 & 14.0 & 16.4 & 24.5 & 24.1 & 16.2 & 25.7 \\
no regional factor & 38.1 & 21.1 & 22.2 & 43.8 & 38.3 & 26.2 & 44.1 \\
\bottomrule
\end{tabular}
\end{table*}

\begin{table*}[tbp]
\caption{Share of hidden reports (\%) inside the calibrated 80\% interval of \auto{}, offline, by distance to the nearest report. The target is 80.}
\label{tab:calib}
\small
\begin{tabular}{lrrrrrrr}
\toprule
Weeks to nearest report & Assam & Meghalaya & Manipur & Nagaland & Tripura & Bihar & Goa \\
\midrule
1 & 81.6 & 81.1 & 81.8 & 80.3 & 80.8 & 81.7 & 83.2 \\
2--3 & 81.5 & 79.4 & 79.9 & 77.5 & 81.6 & 75.7 & 84.6 \\
4--8 & 80.1 & 72.8 & 74.7 & 77.5 & 80.6 & 69.1 & 77.4 \\
9+ & 78.2 & 65.1 & 66.9 & 78.2 & 76.1 & 78.6 & 85.2 \\
\bottomrule
\end{tabular}
\end{table*}

\noindent\begin{minipage}[t]{0.49\linewidth}\vskip0pt\relax
\begin{table}[tbp]
\caption{Offline MAPE (\%) in Assam by distance to the nearest visible report of the same market.}
\label{tab:gaplen}
\small
\begin{tabular}{lrrrr}
\toprule
Method & 1 week & 2--3 weeks & 4--8 weeks & 9+ weeks \\
\midrule
\auto{} & \textbf{10.4} & \textbf{14.9} & \textbf{17.3} & \textbf{24.5} \\
bridge (fixed) & 11.0 & 15.5 & 18.5 & 25.3 \\
no AR bridge & 16.8 & 19.4 & 20.2 & 25.8 \\
no regional factor & 12.8 & 22.3 & 33.8 & 51.2 \\
SAITS & 12.7 & 18.6 & 24.8 & 40.7 \\
chained ML & 15.0 & 21.5 & 27.9 & 39.2 \\
BRITS & 12.4 & 19.2 & 27.1 & 43.3 \\
linear & 11.2 & 19.1 & 30.1 & 60.3 \\
Kalman & 11.9 & 19.8 & 30.7 & 60.2 \\
carry-forward & 20.7 & 28.6 & 35.2 & 61.8 \\
\bottomrule
\end{tabular}
\end{table}

\begin{table}[tbp]
\caption{Runtime to fill one commodity panel in Assam (about 60 target markets and 100 helper series), one CPU core.}
\label{tab:runtime}
\small
\begin{tabular}{lr}
\toprule
Method & Seconds per commodity panel \\
\midrule
carry-forward & 0.06 \\
linear & 0.07 \\
PCHIP & 0.09 \\
SoftImpute & 0.18 \\
Kalman & 0.18 \\
bridge (fixed) & 0.46 \\
kNN & 0.89 \\
\auto{} & 3.89 \\
SAITS & 6.03 \\
DFM-EM & 6.04 \\
BRITS & 18.33 \\
chained ML & 42.06 \\
\bottomrule
\end{tabular}
\end{table}

\end{minipage}\hfill
\begin{minipage}[t]{0.49\linewidth}\vskip0pt\relax
\begin{table}[tbp]
\caption{Offline MAPE (\%) in Assam when 20\% to 80\% of reports are hidden at random.}
\label{tab:missing}
\small
\begin{tabular}{lrrrr}
\toprule
Method & 20\% hidden & 40\% & 60\% & 80\% \\
\midrule
\auto{} & \textbf{9.5} & \textbf{10.4} & \textbf{11.5} & \textbf{13.7} \\
bridge (fixed) & 9.7 & 10.7 & 12.0 & 14.7 \\
PCHIP & 10.7 & 11.8 & 13.7 & 19.2 \\
linear & 10.5 & 11.7 & 13.7 & 19.2 \\
Kalman & 10.7 & 12.2 & 15.3 & 23.8 \\
SoftImpute & 16.1 & 18.0 & 21.3 & 27.6 \\
kNN & 15.9 & 23.1 & 25.5 & 28.7 \\
carry-forward & 15.4 & 17.2 & 22.1 & 30.9 \\
\bottomrule
\end{tabular}
\end{table}

\begin{table}[tbp]
\caption{Offline MAPE (\%) in Assam by commodity (all scenarios), for \auto{} and four baselines. \auto{} has the lowest error among all ten baselines for 17 of 20 commodities.}
\label{tab:commodity}
\footnotesize\setlength{\tabcolsep}{2.7pt}
\begin{tabular}{lrrrrr}
\toprule
Commodity & \auto{} & SAITS & chained ML & linear & carry-fwd. \\
\midrule
Beetroot & 18.3 & 20.0 & 24.0 & \textbf{15.4} & 21.8 \\
Bhindi (Ladies Finger) & \textbf{21.9} & 28.7 & 29.2 & 43.5 & 47.5 \\
Bitter gourd & \textbf{18.5} & 20.7 & 20.7 & 28.3 & 30.8 \\
Bottle gourd & \textbf{18.7} & 32.4 & 31.9 & 24.1 & 25.4 \\
Brinjal & \textbf{25.0} & 40.8 & 36.4 & 37.9 & 42.4 \\
Cabbage & \textbf{20.4} & 32.0 & 30.7 & 48.4 & 62.5 \\
Capsicum & \textbf{20.4} & 26.4 & 37.1 & 34.5 & 36.2 \\
Carrot & \textbf{37.1} & 43.9 & 61.8 & 40.9 & 43.6 \\
Cauliflower & \textbf{23.2} & 49.4 & 50.6 & 53.8 & 56.7 \\
Cucumbar (Kheera) & \textbf{24.1} & 29.0 & 33.1 & 31.7 & 32.1 \\
French Beans (Frasbean) & \textbf{28.2} & 39.7 & 41.7 & 54.4 & 63.6 \\
Garlic & 10.6 & 14.2 & 16.1 & \textbf{9.7} & 14.3 \\
Ginger (Green) & \textbf{18.7} & 22.3 & 21.2 & 22.1 & 23.6 \\
Green Chilli & \textbf{19.2} & 26.6 & 24.2 & 35.6 & 45.5 \\
Lemon & \textbf{20.8} & 34.1 & 35.1 & 85.2 & 87.7 \\
Onion & \textbf{11.7} & 23.3 & 26.6 & 26.7 & 27.5 \\
Potato & \textbf{17.7} & 17.9 & 18.8 & 22.3 & 22.4 \\
Pumpkin & \textbf{15.6} & 22.8 & 21.7 & 25.2 & 27.0 \\
Raddish & 24.5 & \textbf{21.2} & 28.3 & 22.9 & 26.3 \\
Tomato & \textbf{17.9} & 37.9 & 37.3 & 59.3 & 61.1 \\
\bottomrule
\end{tabular}
\end{table}

\end{minipage}
\clearpage
\section{Supplementary Visual Results}\label{app:visuals}
These plots provide supporting views of the saved experiments, without additional model fitting. The filtered-state comparison uses full-panel parameter estimates, not strictly past-only fits.

\newenvironment{appendixflowfigure}{%
  \par\addvspace{7pt}\noindent\begin{minipage}{\linewidth}%
  \captionsetup{type=figure}\centering
}{\end{minipage}\par\addvspace{5pt}}
\renewenvironment{figure}[1][]{\begin{appendixflowfigure}}{\end{appendixflowfigure}}

\noindent\begin{minipage}[t]{0.49\linewidth}\vskip0pt\relax
\begin{figure}
\includegraphics[width=\linewidth]{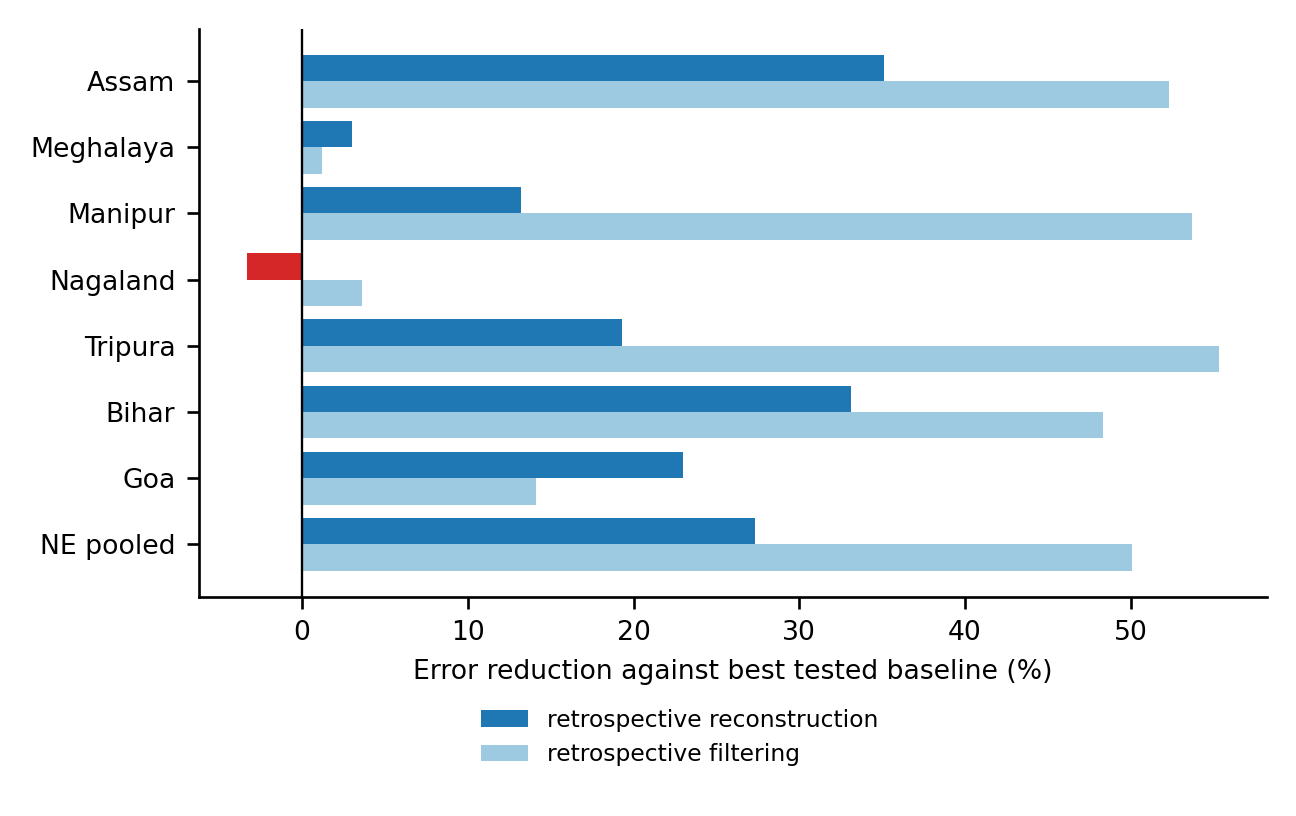}
\caption{Relative error reduction over the best tested baseline in each state. The filtered-state series uses retrospective full-panel fits; it is not a nowcasting result. Negative values indicate higher error.}
\Description{Paired horizontal bars show retrospective reconstruction and filtered-state relative error reductions across states and the North-East pool.}
\label{fig:gain}
\end{figure}
\end{minipage}\hfill
\begin{minipage}[t]{0.49\linewidth}\vskip0pt\relax
\begin{figure}
\includegraphics[width=\linewidth]{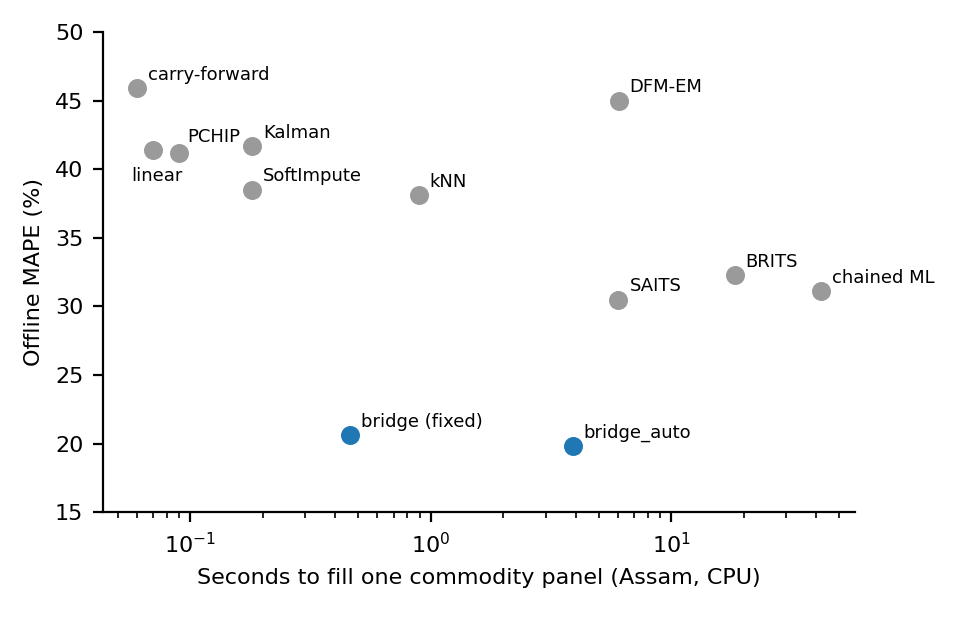}
\caption{Assam retrospective MAPE against measured fill time per commodity panel (one CPU core; logarithmic time axis). These are configuration-specific measurements, not model-family complexity bounds.}
\Description{Scatter plot of error against runtime for the twelve tested implementations, with fixed and adaptive bridges highlighted.}
\label{fig:cost}
\end{figure}
\end{minipage}

\noindent\begin{minipage}[t]{0.49\linewidth}\vskip0pt\relax
\begin{figure}
\includegraphics[width=0.94\linewidth]{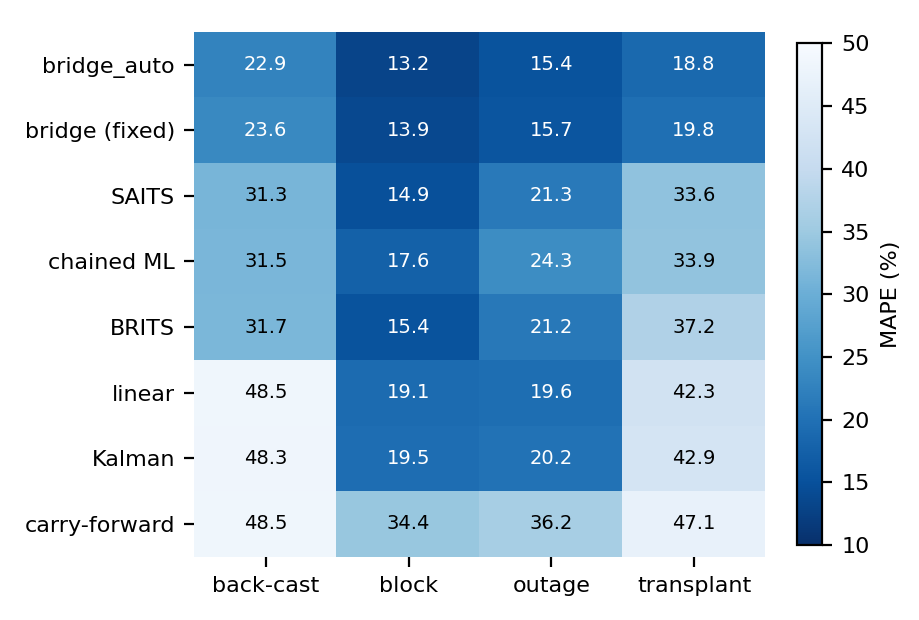}
\caption{Assam retrospective MAPE by saved scenario. Block and outage entries correspond to the displayed heavy-method comparison runs, not accuracy in genuinely unreported weeks.}
\Description{Heat map of errors for eight implementations across back-cast, block, outage and transplant scenarios.}
\label{fig:scenarios}
\end{figure}
\end{minipage}\hfill
\begin{minipage}[t]{0.49\linewidth}\vskip0pt\relax
\begin{figure}
\includegraphics[width=\linewidth]{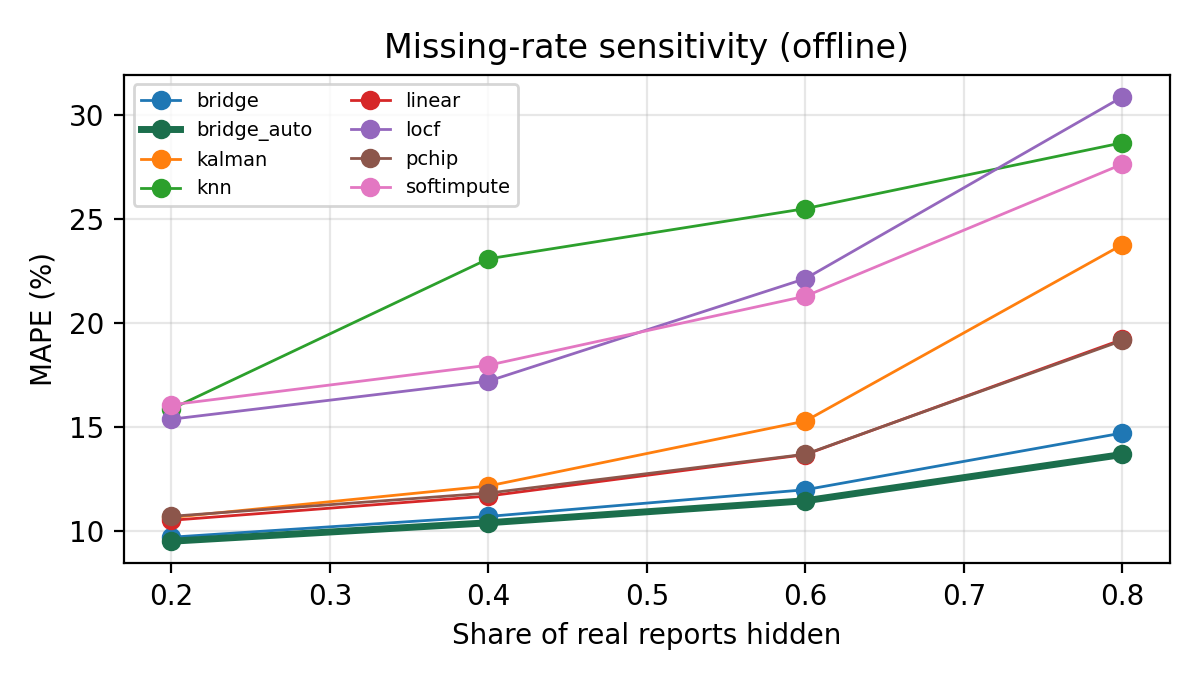}
\caption{Assam error under additional random deletion of reported values. The 20--80\% masking sweep is distinct from the naturally occurring missingness mechanism.}
\Description{Error curves for selected methods as the proportion of deliberately hidden observed reports increases.}
\label{fig:missing}
\end{figure}
\end{minipage}

\endgroup
\end{document}